\documentclass[preprint,authoryear]{elsarticle}
\journal{Expert Systems With Applications}
\usepackage{cite}
\usepackage{graphicx}
\usepackage{amsmath}
\usepackage[caption=false,font=footnotesize]{subfig}
\usepackage{url}
\usepackage[strings]{underscore} 
\usepackage{amssymb}  
\usepackage{ctable}
\usepackage[ruled,vlined,linesnumbered]{algorithm2e}
\usepackage{makecell} 
\usepackage{xcolor} 

\begin{document}

\title{COVID-19 Detection on Chest X-Ray Images: A comparison of CNN architectures and ensembles}

\author{Fabricio Aparecido Breve}
\ead{fabricio.breve@unesp.br}
\tnotetext[t1]{© 2022. This manuscript version is made available under the CC-BY-NC-ND 4.0 license.}

\address{Institute of Geosciences and Exact Sciences,
S\~{a}o Paulo State University (UNESP) J\'{u}lio de Mesquita Filho, Rio Claro, SP, 13506-900, Brazil}

\newpageafter{author}

\begin{abstract}

COVID-19 quickly became a global pandemic after only four months of its first detection. It is crucial to detect this disease as soon as possible to decrease its spread. The use of chest X-ray (CXR) images became an effective screening strategy, complementary to the reverse transcription-polymerase chain reaction (RT-PCR). Convolutional neural networks (CNNs) are often used for automatic image classification and they can be very useful in CXR diagnostics. In this paper, 21 different CNN architectures are tested and compared in the task of identifying COVID-19 in CXR images. They were applied to the COVIDx8B dataset, a large COVID-19 dataset with $16,352$ CXR images coming from patients of at least $51$ countries. Ensembles of CNNs were also employed and they showed better efficacy than individual instances. The best individual CNN instance results were achieved by DenseNet169, with an accuracy of 98.15\% and an F1 score of 98.12\%. These were further increased to 99.25\% and 99.24\%, respectively, through an ensemble with five instances of DenseNet169. These results are higher than those obtained in recent works using the same dataset.

\end{abstract}


\maketitle

\section{Introduction}

COVID-19 is an infectious disease caused by the Severe Acute Respiratory Syndrome CoronaVirus 2 (SARS-CoV-2) \citep{Khan2021}. It quickly became a global pandemic in less than four months after its first detection in December 2019 in Wuhan, China \citep{Monshi2021}. As of February 2022, it has over 434 million confirmed cases and almost 6 million deaths reported to World Health Organization \citep{WHO2022}. Early detection of positive COVID-19 cases is critical for avoiding the virus's spread.

The most common technique for diagnosing COVID-19 is known as transcriptase-polymerase chain reaction (RT-PCR). It detects SARS-CoV-2 through collected respiratory specimens of nasopharyngeal or oropharyngeal swabs. However, RT-PCR testing is expensive, time-consuming, and shows poor sensitivity \citep{Monshi2021, Mostafiz2020}, especially in the first days of exposure to the virus \citep{Long2020}. Up to 54\% of COVID-19 patients may have an initial negative RT-PCR result \citep{Arevalo2020}.

Patients that receive a false negative diagnosis may contact and infect other people before they are tested again. Therefore, it is important to have alternative methods to detect the disease, such as Chest X-ray (CXR) images. CXR equipment is widely available in hospitals and CXR images are cheap and fast to acquire. They can be inspected by radiologists to find visual indicators of the virus \citep{Feng2020}.

In the past decade, the rise of deep learning methods \citep{Lecun2015, Goodfellow2016, Schmidhuber2015}, especially the convolutional neural networks (CNNs), were responsible for many advances in automatic image classification \citep{Krizhevsky2012}. CXR diagnostic using deep learning methods is a mechanism that can be explored to surpass the limitations of RT-PCR insufficient test kits, waiting time of test results, and test costs \citep{Mostafiz2020}.

Many studies concerning the application of CNNs to COVID-19 diagnostic on CXR images were published since the last year \citep{Khan2021,Nigam2021,Ismael2021,Abbas2021,Hira2021,Alawad2021,Narin2021,Monshi2021,Heidari2020,Jia2021,Karthik2021,Mostafiz2020,Shorfuzzaman2020,Chhikara2021}. However, most of them used relatively small and more homogeneous datasets. In this paper, the COVIDx8B dataset\footnote{Scripts to build the COVIDx8B dataset are available at \url{https://github.com/lindawangg/COVID-Net/blob/master/docs/COVIDx.md}} \citep{Zhao2021} is used. It has $16,352$ CXR images, from which $2,358$ are COVID-19 positive and the remaining are from both healthy and pneumonia patients. Released in March 2021, this dataset is composed of images from six other open-source chest radiography datasets. Therefore it is larger and more heterogeneous than earlier available datasets. However, there are only a few works that used this dataset so far \citep{Pavlova2021, Zhao2021, Dominik2021}. A recent survey on applications of artificial intelligence in the COVID-19 pandemic \citep{Khan2021} reviewed dozens of papers, including $16$ papers on CNNs applied to CXR images and all of them used earlier available datasets which are smaller than COVIDx8B.

In this paper, a comparison of $21$ different CNN models applied to the COVIDx8B dataset is presented, including popular architectures such as VGG \citep{Simonyan2015}, ResNet \citep{He2016}, DenseNet \citep{Huang2017}, and EfficientNet \citep{Tan2019}. They were all trained in the same conditions with the training and test subsets defined by the dataset authors. The initial weights of all methods were defined to those trained on the ImageNet dataset \citep{ILSVRC15}, which is commonly used in transfer learning scenarios \citep{Oquab2014}. The accuracy, sensitivity (TPR), precision (PPV), and F1 score were evaluated using the test subset. Later, some models' continuous output (before the classification layer) were combined (ensembles) to overcome individual limitations and provide better classification results.

The remainder of this paper is organized as follows. Section~\ref{sec:RelatedWork} shows related work, in which CNNs were used to detect COVID-19 on CXR images. Section~\ref{sec:Dataset} presents the COVIDx8B dataset. Section~\ref{sec:CNNArchitectures} shows the CNN architectures employed in this paper. Section~\ref{sec:CNNComparison} shows the computer simulations comparing the proposed models and other recent approaches from the literature for COVID-19 classification on CXR images using the same dataset. Section~\ref{sec:CNNEnsembles} shows the computer simulations with CNN ensembles, improving the classification performance of individual models. Finally, the conclusions are drawn in Section~\ref{sec:Conclusions}.

\section{Related Work}
\label{sec:RelatedWork}

Many studies have investigated the use of machine learning techniques to detect COVID-19. Many of the researchers used CNN techniques and CXR images and faced challenges due to the lack of available datasets \citep{Alawad2021}. While many authors provided tables comparing results achieved in different works, the comparisons are not fair, since the used datasets are frequently different and pose different levels of challenge. Therefore, here the related works are described focusing on what architectures have been used to handle the problem of COVID-19 detection on CXR images and the size of the evaluated datasets.

\citet{Nigam2021} used VGG16, DenseNet121, Xception, NASNet, and EfficientNet in a dataset with 16,634 images. Though this dataset is slightly larger than COVIDx8B, unfortunately, the authors did not make it publicly available. The highest accuracy was $93.48\%$ obtained with EfficientNetB7.

\citet{Ismael2021} used ResNet18, ResNet50, ResNet101, VGG16, and VGG19 for deep feature extraction and support vector machines (SVM) for CXR images classification. The highest accuracy was $94.7\%$ obtained with ResNet50. However, they used a small dataset with only $380$ CXR images.

\citet{Abbas2021} validated a deep CNN called Decompose, Transfer, and Compose (DeTraC) for COVID-19 CXR images classification with $93.1\%$ accuracy. They used a combination of two small datasets, totaling $196$ images.

\citet{Hira2021} used the AlexNet, GoogleNet, ResNet-50, Se-ResNet-50, DenseNet121, Inception V4, Inception ResNet V2, ResNeXt-50, and Se-ResNeXt-50 architectures. Se-ResNeXt-50 achieved the highest classification accuracy of $99.32\%$. They used a combination of four datasets, totalling $8,830$ CXR images.

\citet{Alawad2021} used VGG16 both as a stand-alone classifier and as a feature extractor for SVM, Random-Forests (RF), and Extreme-Gradient-Boosting (XGBoost) classifiers. VGG-16 and VGG16+SVM models provide the best performance with $99.82\%$ accuracy. They used a combination of five datasets, totaling $7,329$ CXR images.

\citet{Narin2021} used ResNet50, ResNet101, ResNet152, InceptionV3, and Inception-ResNetV2. ResNet50 achieved the highest classification performance with $96.1\%$, $99.5\%$, and $99.7\%$ accuracy on three different datasets, totalling $7,406$ CXR images.

\citet{Monshi2021} focused on data augmentation and CNN hyperparameters optimization, increasing VGG19 and ResNet50 accuracy. They also proposed CovidXrayNet, a model based on EfficientNet-B0, which achieved an accuracy of $95.82\%$ on an earlier version of the COVIDx dataset with $15,496$ CXR images.

\citet{Heidari2020} focused on preprocessing algorithms to improve the performance of VGG16. They used a dataset with $8,474$ CXR images and reached $94.5\%$ accuracy.

\citet{Jia2021} proposed a modified MobileNet to classify CXR and CT images. They applied their method to a CXR dataset with $7,592$ CXR images and achieved $99.3\%$ accuracy. They also applied it to an earlier version of COVIDx with $13,975$ CXR images, achieving $95.0\%$ accuracy.

\citet{Karthik2021} proposed a custom CNN architecture which they called Channel-Shuffled Dual-Branched (CSDB). They achieved an accuracy of $99.80\%$ on a combination of seven datasets, totaling $15,265$ images.

\citet{Mostafiz2020} used a hybridization of CNN (ResNet50) and discrete wavelet transform (DWT) features. The random forest-based bagging approach was used for classification. They combined different datasets and used data augmentation techniques to produce a total of $4809$ CXR images and achieved $98.5\%$ accuracy.

\citet{Shorfuzzaman2020} used VGG16, ResNet50V2, Xception, MobileNet, and DenseNet121 in a transfer learning scenario. They collected CXR images from different sources to compose a dataset with $678$ images. The best accuracy ($98.15\%$) was achieved with ResNet50V2. They also made an ensemble of the four best models (ResNet50V2, Xception, MobileNet, and DenseNet121) with the final output obtained by majority voting, raising the accuracy to $99.26\%$.

\citet{Chhikara2021} proposed a InceptionV3 based-model and applied it to three different datasets with $11,244$, $8,246$, and $14,486$ CXR images, respectively. The model has reached an accuracy of $97.7\%$, $84.95\%$, and $97.03\%$ on the mentioned datasets, respectively.

\citet{Pavlova2021} proposed the COVIDx8B dataset, which they claim is the largest and most diverse COVID-19 CXR dataset in open access form, and the COVID-Net CXR-2 model, a CNN specially tailored for COVID-19 detection on CXR images using machine-driven design, which achieved an accuracy of $95.5\%$.

\citet{Zhao2021} used ResNet50V2 to classify the COVIDx8B dataset with an accuracy of $96.5\%$ in the best scenario.

\citet{Dominik2021} proposed a lightweight architecture called BaseNet and achieved an accuracy of $95.50\%$ on COVIDx8B. He also used an ensemble composed of BaseNet, VGG16, VGG19, ResNet50, DenseNet121, and Xception to achieve $97.75\%$ accuracy. It was further increased to $99.25\%$ using an optimal classification threshold.

\section{Dataset}
\label{sec:Dataset}

\begin{figure}
  \centering
   \setlength\tabcolsep{1.5pt}
   \subfloat{\includegraphics[height=3.02cm]{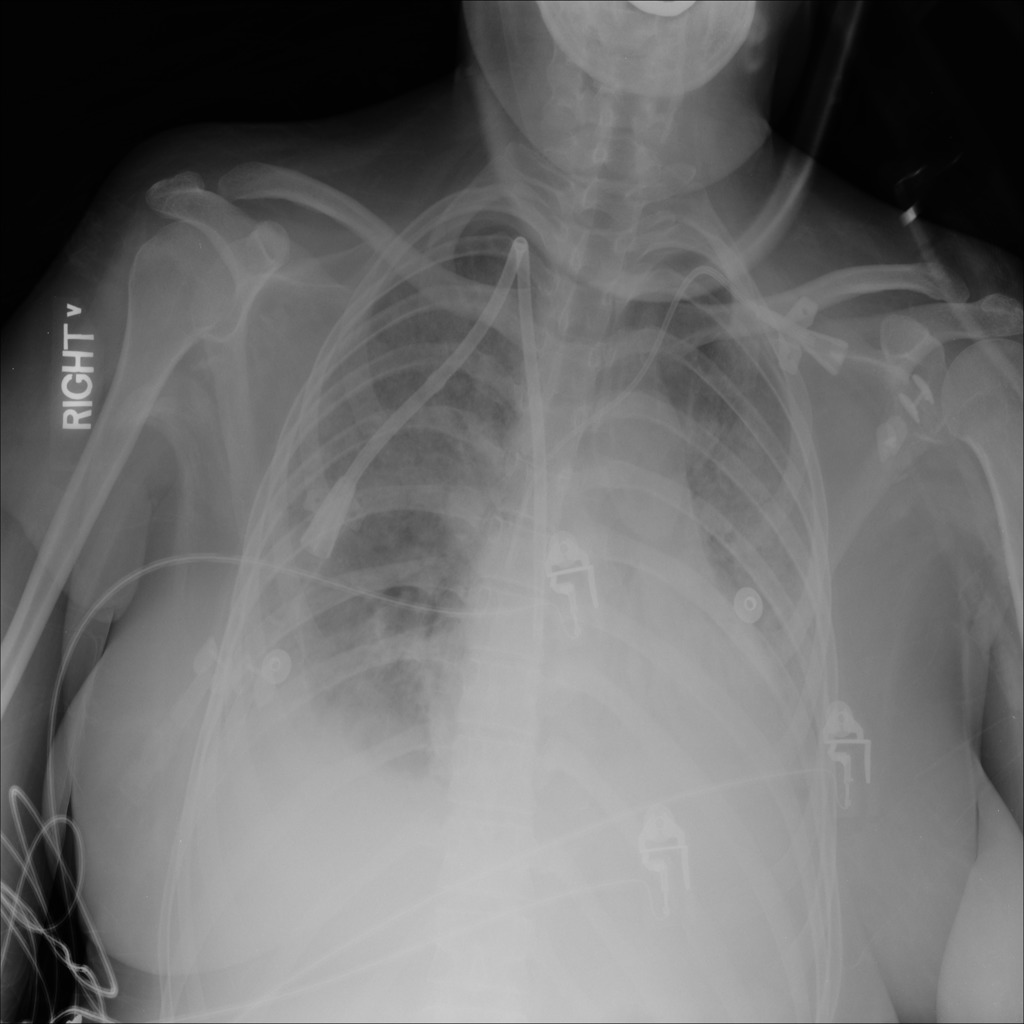}}
   \subfloat{\includegraphics[height=3.02cm]{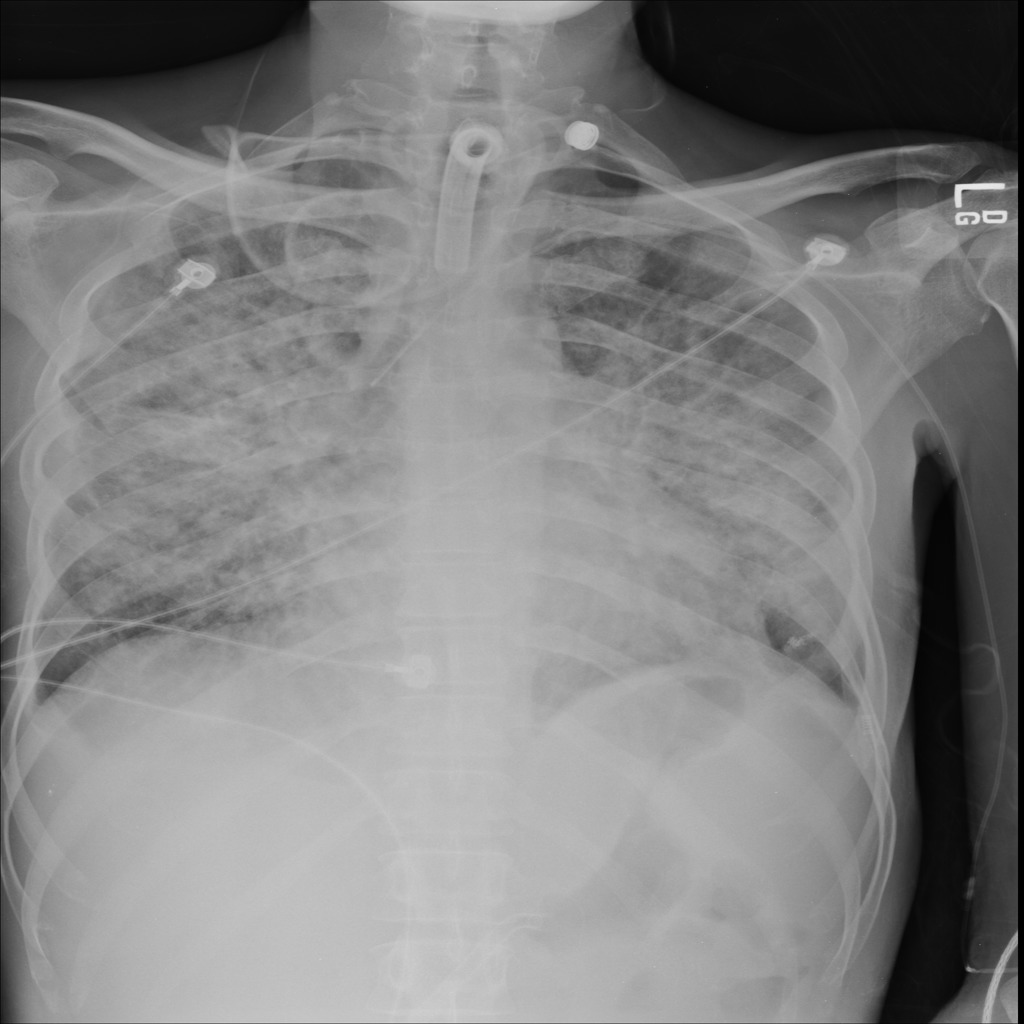}}
   \subfloat{\includegraphics[height=3.02cm]{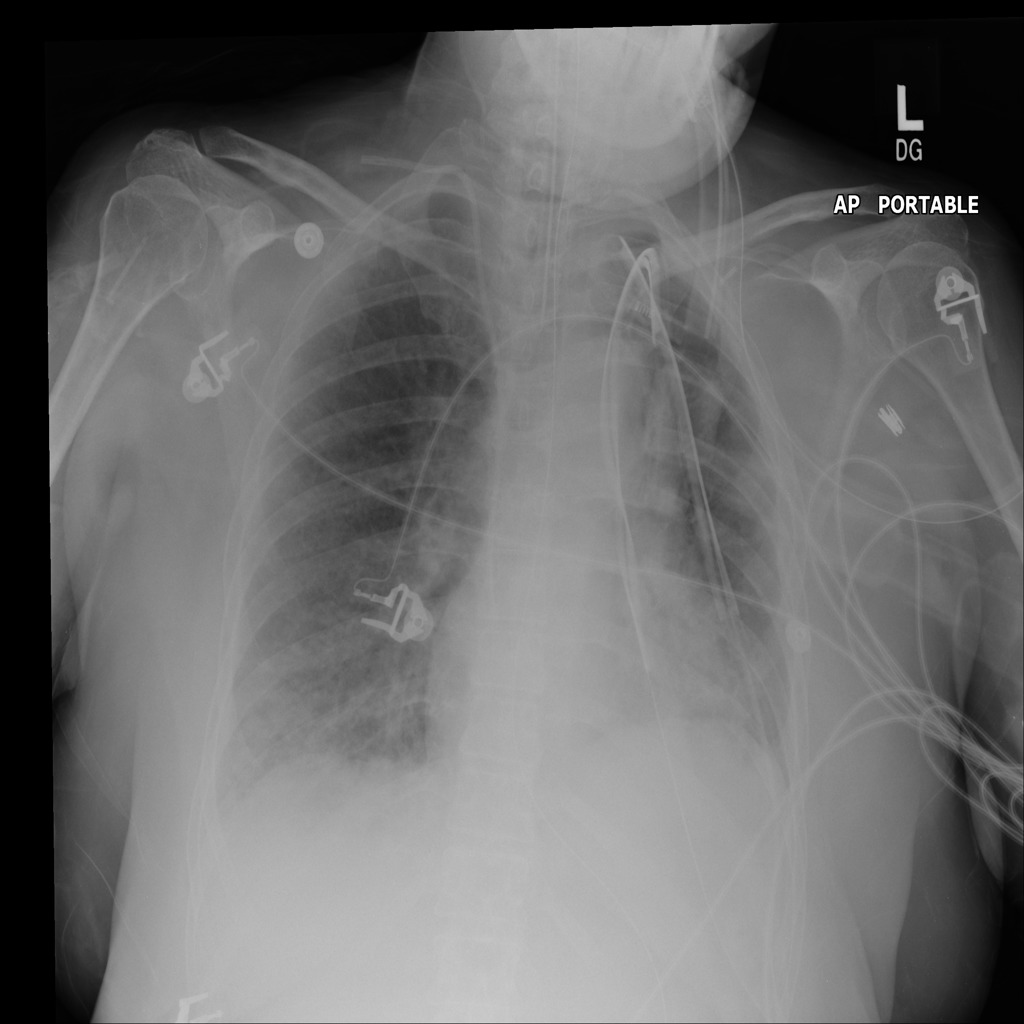}}
   \subfloat{\includegraphics[height=3.02cm]{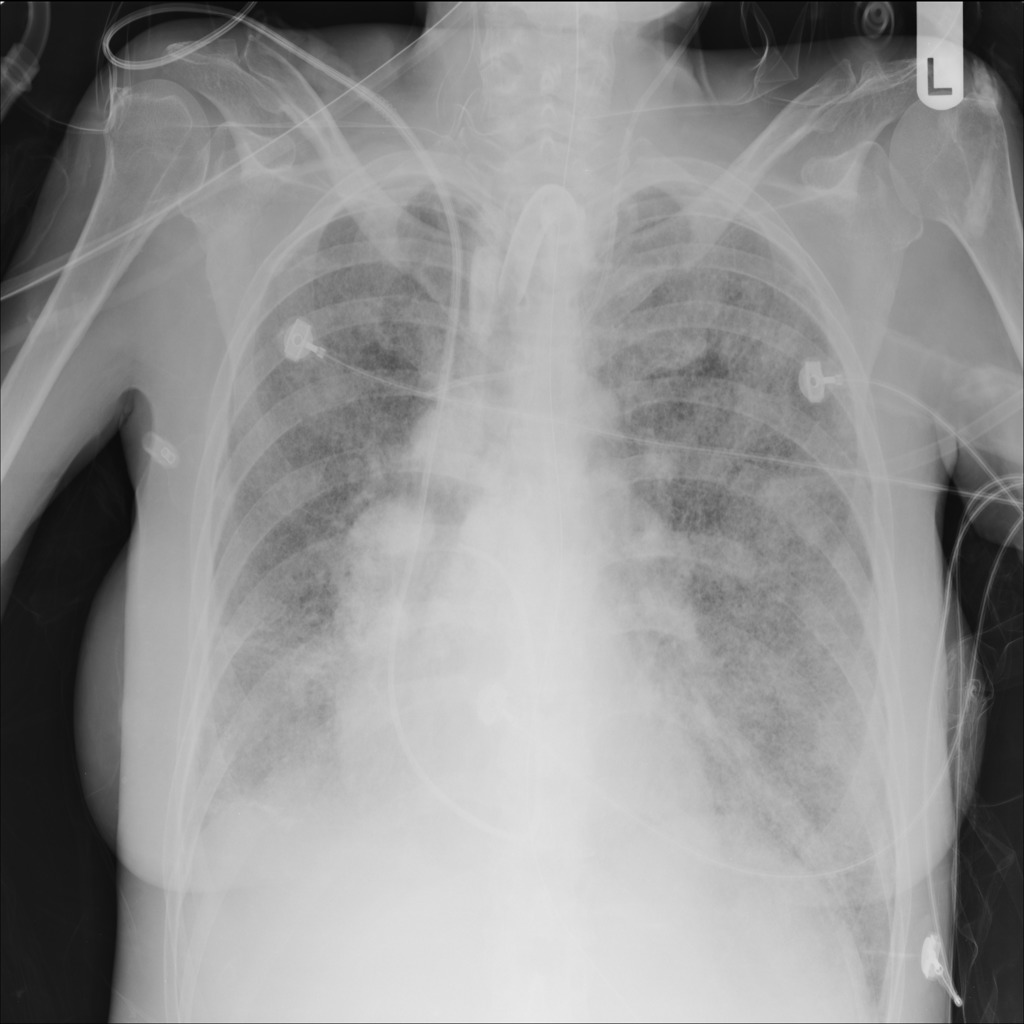}} \\
   \subfloat{\includegraphics[height=2.8cm]{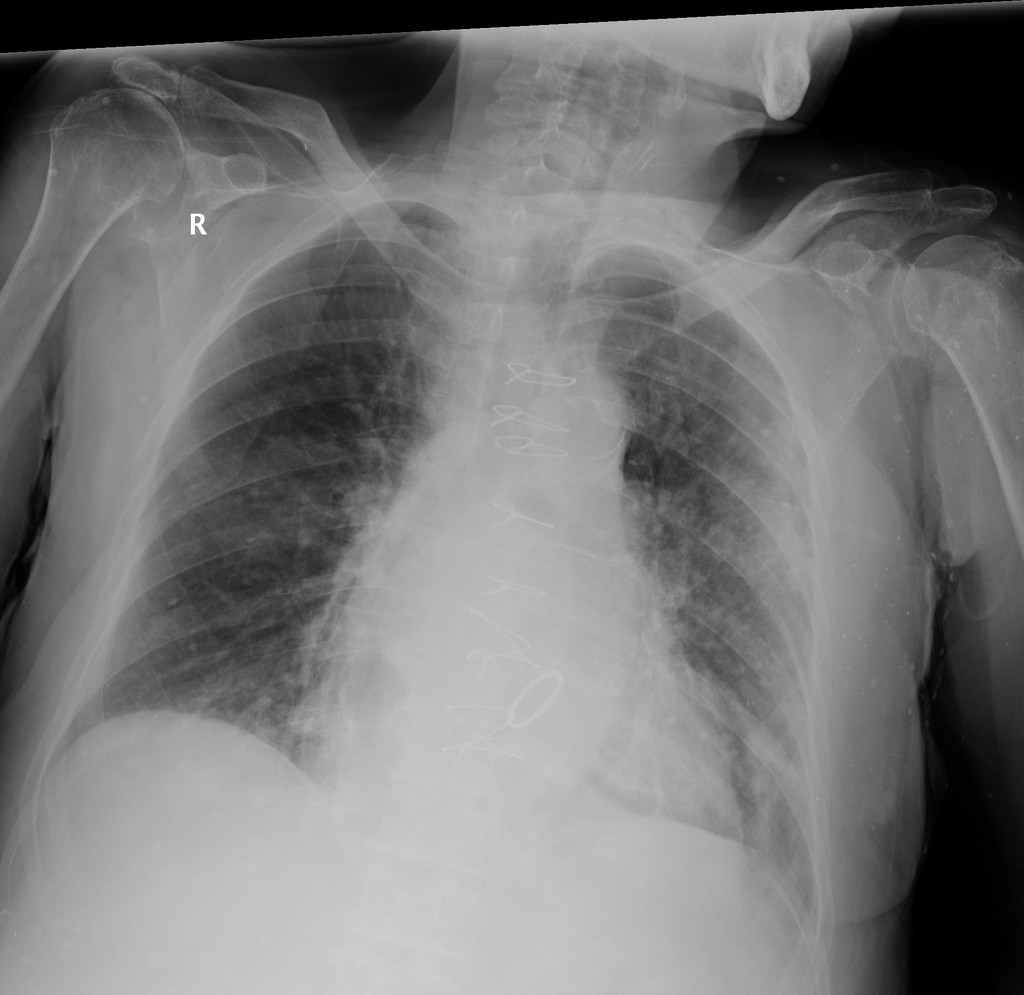}}
   \subfloat{\includegraphics[height=2.8cm]{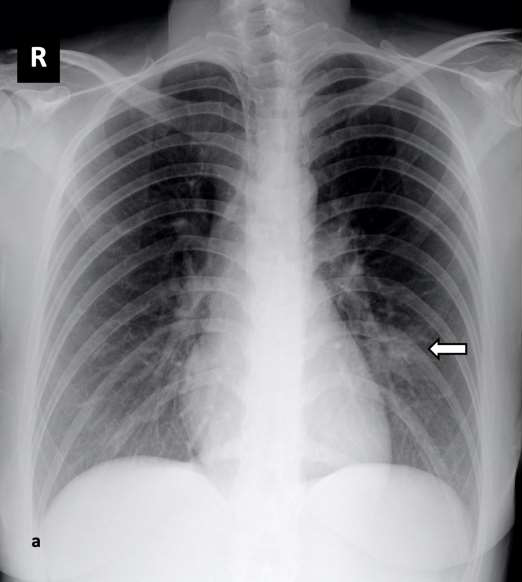}}
   \subfloat{\includegraphics[height=2.8cm]{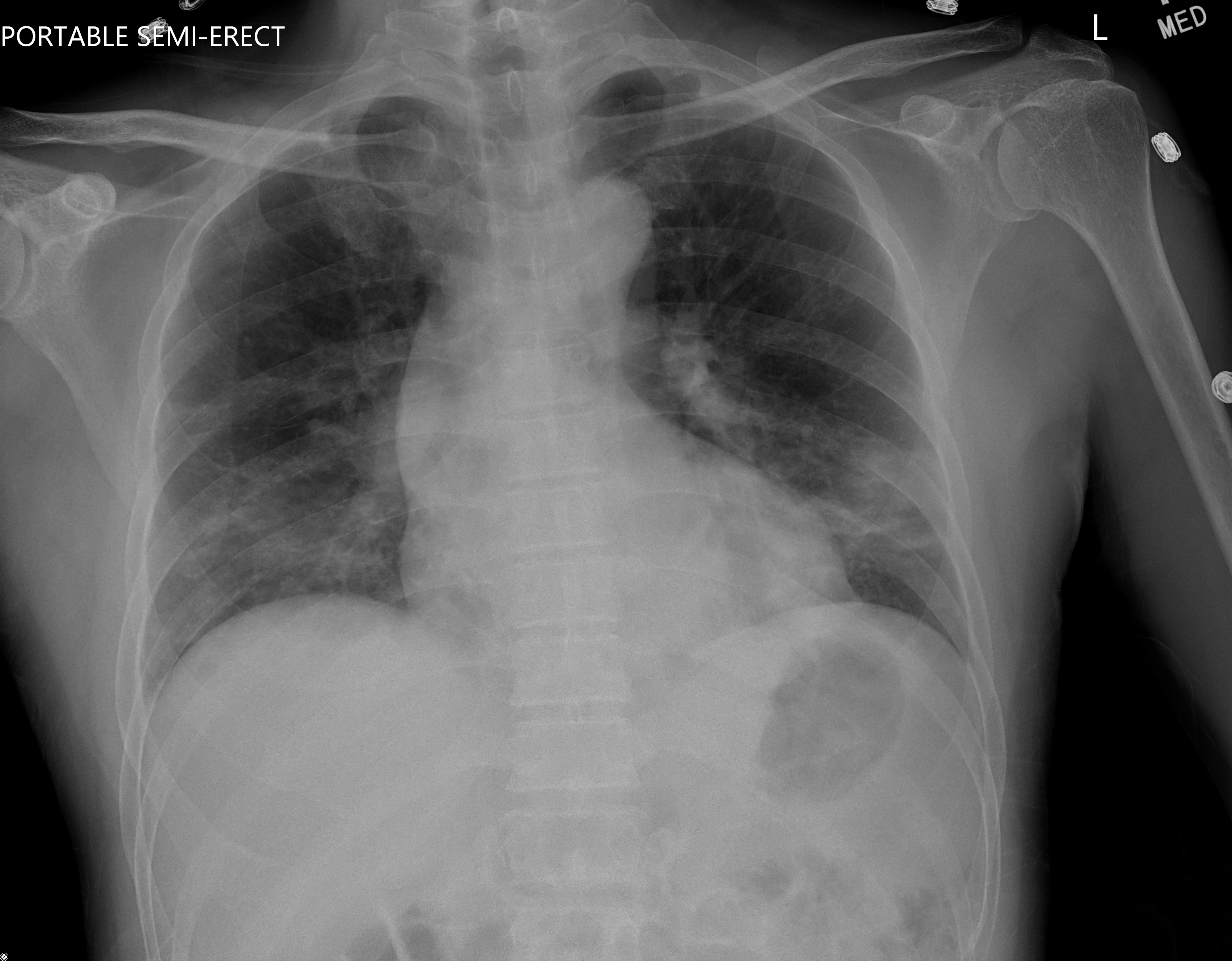}}
   \subfloat{\includegraphics[height=2.8cm]{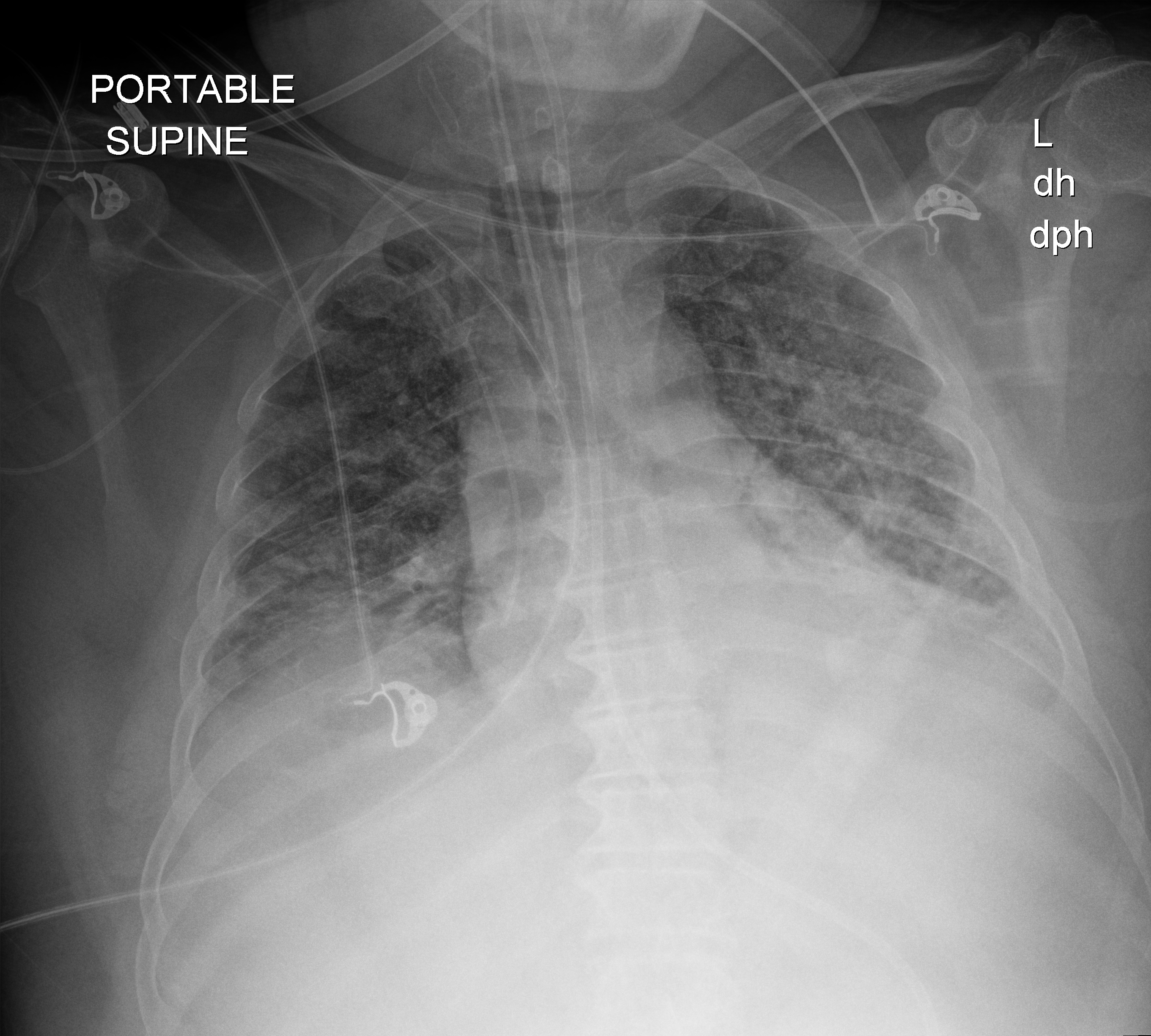}} \\
  \caption{Examples of CXR images from the COVIDx8B dataset. The first row shows COVID-19 negative patient cases, and the second row shows COVID-19 positive patient cases.}
  \label{fig:Dataset}
\end{figure}

Most of the early research regarding COVID-19 detection on CXR images suffered from the lack of available datasets \citep{Alawad2021}. The authors would frequently combine different smaller datasets, so fairly comparing the results was impossible. COVIDx8B is a large and heterogeneous COVID-19 CXR benchmark dataset with $16,352$ CXR images coming from patients of at least $51$ countries \citep{Pavlova2021}. It is constructed with images extracted from six open-source chest radiography datasets, which are shown in Table~\ref{tab:Dataset}. Notice that the sum of the images in the source datasets is much larger than the size of COVIDx8B since not all of them were selected by the authors. Example images from the COVIDx8B dataset are shown in Figure~\ref{fig:Dataset}.

\begin{table}

\resizebox{\textwidth}{!}{%

\begin{tabular}{p{8cm}cp{4cm}}

\toprule

{\bf Source dataset} & {\bf Size} & {\bf Reference} \\

\toprule

covid-chestxray-dataset &        950 & \citet{Cohen2020} \\

\midrule

Figure 1 COVID-19 Chest X-ray Dataset Initiative &         55 & \citet{Chung2020a} \\

\midrule

Actualmed COVID-19 Chest X-ray Dataset Initiative &        238 & \citet{Chung2020b} \\

\midrule

COVID-19 Radiography Database - Version 3 &     21,165 & \citet{Chowdhury2020,Rahman2021} \\

\midrule

RSNA Pneumonia Detection Challenge &     29,684 & \citet{Wang2017} \\

\midrule

RSNA International COVID-19 Open Radiology Database (RICORD) &      1,257 & \citet{Tsai2021} \\

\bottomrule

\end{tabular}

}

\caption{List of datasets that compose the COVIDx8B benchmark dataset.}
\label{tab:Dataset}

\end{table}

Though COVIDx8B does not include information on patients' demographics, half of their source datasets do. The covid-chestxray-dataset has $559$ registers from male patients and $311$ registers from female patients. The average age is $54$ years old. The COVID-19 positive registers are from $346$ male and $175$ female patients, with an average age of $56$ years old.
The Figure 1 COVID-19 Chest X-ray Dataset Initiative has only $55$ registers, most of them do not indicate sex. Among the remaining, there are $11$ male patients and $11$ female patients. Only $21$ patients have their exact age registered and the average is $52$ years old. All patients with the exact age described are COVID-19 positive or unlabeled. The RSNA International COVID-19 Open Radiology Database (RICORD) only has COVID-19 positive cases. They come from $645$ male and $353$ female patients, with an average age of $56$ years old.

Four of the source datasets have both COVID-19 positive and negative cases. The RSNA Pneumonia Detection Challenge has only COVID-19 negative cases (non-COVID pneumonia, normal, etc.) and The RSNA International COVID-19 Open Radiology Database (RICORD) has only COVID-19 positive cases.

COVID8xB training subset is composed of $15,952$ images, from which $2,158$ are COVID-19 positive and $13,794$ are COVID-19 negative. The negative group also includes images of patients with non-COVID-19 pneumonia, which poses a major challenge as it is usually difficult to distinguish between COVID-19 and non-COVID19 pneumonia. The test subset has $200$ COVID-19 positive images from $178$ different patients and $200$ COVID-19 negative images. In the negative group, $100$ images are from healthy patients. The other $100$ images are from non-COVID pneumonia patients. The test images were randomly selected from international patient groups curated by the Radiological Society of North America (RSNA) \citep{Wang2017,Tsai2021}. The images were annotated by an international group of scientists and radiologists from different institutes around the world. The test set was selected in such a way to ensure no patient overlap between training and test sets \citep{Pavlova2021}.

\section{CNN Architectures}
\label{sec:CNNArchitectures}

This section presents the CNN architectures explored in this work. It also describes the layers added to complete the models and perform the CXR images classification. Table~\ref{tab:Models} shows the $21$ tested architectures, some of their characteristics, and their respective literature references.

\begin{table}

\resizebox{\textwidth}{!}{%

\begin{tabular}{rcccl}

\toprule

{\bf Model} & {\bf \makecell{Input Image \\ Resolution}} & {\bf \makecell{Output of Last \\ Conv. Layer}} & {\bf \makecell{Trainable \\ Parameters}} & {\bf Reference} \\

\toprule

DenseNet121 &  $224 \times 224$ & $7 \times 7 \times 1024$ &  $7,216,770$ & \citet{Huang2017} \\

DenseNet169 &  $224 \times 224$ & $7 \times 7 \times 1664$ & $12,911,234$ & \citet{Huang2017} \\

DenseNet201 &  $224 \times 224$ & $7 \times 7 \times 1920$ & $18,585,218$ & \citet{Huang2017} \\

EfficientNetB0 &  $224 \times 224$ & $7 \times 7 \times 1280$ &  $4,335,998$ & \citet{Tan2019} \\

EfficientNetB1 &  $240 \times 240$ & $8 \times 8 \times 1280$ &  $6,841,634$ & \citet{Tan2019} \\

EfficientNetB2 &  $260 \times 260$ & $9 \times 9 \times 1408$ &  $8,062,212$ & \citet{Tan2019} \\

EfficientNetB3 &  $300 \times 300$ & $10 \times 10 \times 1536$ & $11,090,218$ & \citet{Tan2019} \\

InceptionResNetV2 &  $299 \times 299$ & $8 \times 8 \times 1536$ & $54,670,178$ & \citet{Szegedy2017} \\

InceptionV3 &  $299 \times 299$ & $8 \times 8 \times 2048$ & $22,293,410$ & \citet{Szegedy2016} \\

 MobileNet &  $224 \times 224$ & $7 \times 7 \times 1024$ &  $3,469,890$ & \citet{Howard2017} \\

MobileNetV2 &  $224 \times 224$ & $7 \times 7 \times 1280$ &  $2,552,322$ & \citet{Sandler2018} \\

NASNetMobile &  $224 \times 224$ & $7 \times 7 \times 1056$ &  $4,504,084$ & \citet{Zoph2018} \\

 ResNet101 &  $224 \times 224$ & $7 \times 7 \times 2048$ & $43,077,890$ & \citet{He2016} \\

ResNet101V2 &  $224 \times 224$ & $7 \times 7 \times 2048$ & $43,053,954$ & \citet{He2016ResNetV2} \\

 ResNet152 &  $224 \times 224$ & $7 \times 7 \times 2048$ & $58,744,578$ & \citet{He2016} \\

ResNet152V2 &  $224 \times 224$ & $7 \times 7 \times 2048$ & $58,712,962$ & \citet{He2016ResNetV2} \\

  ResNet50 &  $224 \times 224$ & $7 \times 7 \times 2048$ & $24,059,650$ & \citet{He2016} \\

ResNet50V2 &  $224 \times 224$ & $7 \times 7 \times 2048$ & $24,044,418$ & \citet{He2016ResNetV2} \\

     VGG16 &  $224 \times 224$ & $7 \times 7 \times 512$ & $14,846,530$ & \citet{Simonyan2015} \\

     VGG19 &  $224 \times 224$ & $7 \times 7 \times 512$ & $20,156,226$ & \citet{Simonyan2015} \\

  Xception &  $299 \times 299$ & $10 \times 10 \times 2048$ & $21,332,010$ & \citet{Chollet2017} \\

\bottomrule

\end{tabular}

}

\caption{CNN architectures, some of their characteristics, and their references.}
\label{tab:Models}

\end{table}

\begin{figure}
    \centering
    \includegraphics[height=7cm]{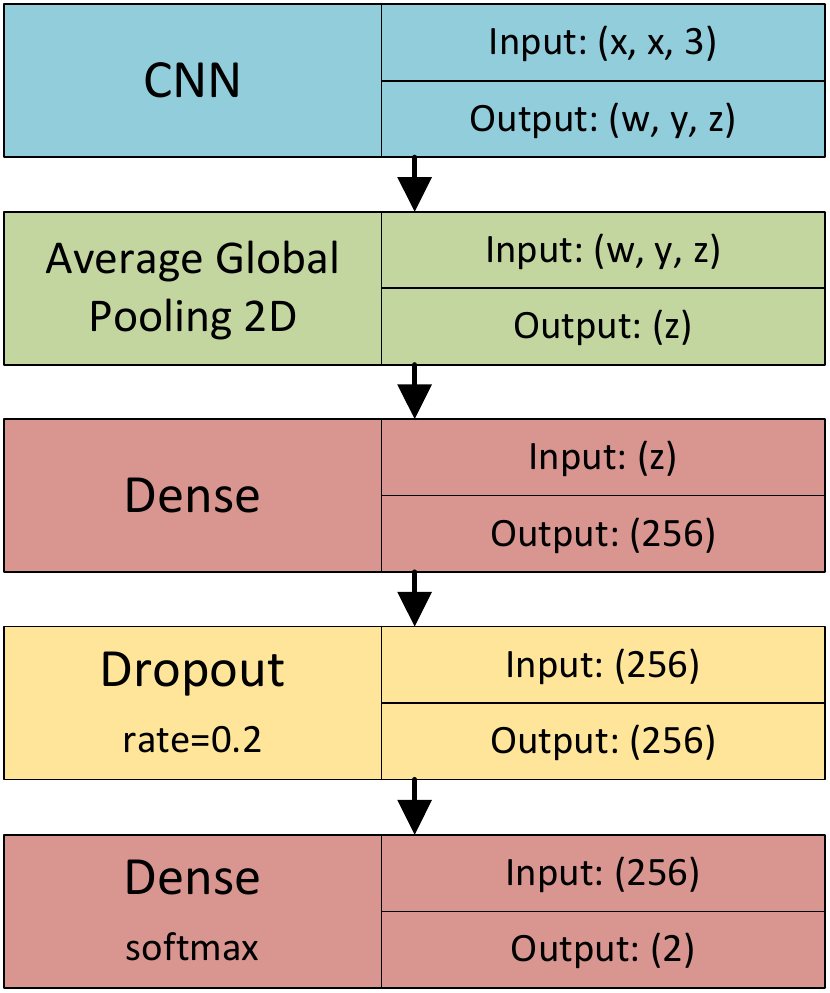}
    \caption{The proposed CNN Transfer Learning architecture.}
    \label{fig:CNN-Diagram}
\end{figure}

The output of the last convolutional layer of the original CNN is fed to a global average pooling layer. Following, there is a dense layer with $256$ neurons using ReLU (Rectified Linear Unit) activation function, a dropout layer with a $20\%$ rate, and a softmax classification layer. This proposed architecture is illustrated in Figure~\ref{fig:CNN-Diagram}, where $x$ indicates the horizontal and vertical input size of the CNN (image size), while $w$, $y$, and $z$ indicate the size of the CNN output in its last convolutional layer. These values depend on the original CNN architecture and they are indicated in Table~\ref{tab:Models}. The table also shows the number of trainable parameters in each CNN architecture, including both their original layers and the dense layers added for COVID-19 classification.

\section{CNN Comparison}
\label{sec:CNNComparison}

In this section, the computer simulations comparing the CNN models applied to the COVIDx8B dataset are presented. All simulations were performed using Python and TensorFlow in three desktop computers with NVIDIA GeForce GPU boards: GTX 970, GTX 1080, and RTX 2060 SUPER, respectively\footnote{The source code is available at \url{https://github.com/fbreve/covid-cnn}}.

No pre-processing was applied, except for those steps pre-defined by each CNN architecture, which is basically resizing the image to the CNN input size and normalizing the input range. In all tested scenarios, each CNN had its weights initially set to those pre-trained on the Imagenet dataset \citep{ILSVRC15}, which has millions of images and hundreds of classes. This dataset is frequently used in transfer learning scenarios.

The training phase was conducted using the Adam optimizer \citep{Kingma2014}. The learning rate was set to $10^{-5}$ for the original CNN layers and $10^{-3}$ for the dense layers proposed in this work. The idea is to allow bigger weight changes in the classification layers, which need to be trained from scratch, while only fine-tuning the CNN layers, taking advantage of the weights previously learned from the Imagenet dataset.

From the training subset, $20\%$ of the images are randomly taken to compose the validation subset, using stratification to keep the same classes proportion. Since the training subset is unbalanced, different class weights were defined for each class: $0.5782$ and $3.6960$ for negative and positive classes, respectively. These values were calculated based on TensorFlow documentation\footnote{TensorFlow documentation on class weights is available at \url{https://www.tensorflow.org/tutorials/structured_data/imbalanced_data}}:
\begin{equation}
  w_i = \frac{1}{c_i} \times \frac{t}{2}
\end{equation}
where $w_i$ is the class $i$ weight, $c_i$ is the amount of examples belonging to class $i$, and $t$ is the total amount of examples.

All models were trained for up to $50$ epochs. An early stopping criterion was set to interrupt the training phase if the loss on the validation set did not decrease during the last $10$ epochs. The final weights are always restored to those adjusted in the epoch that achieved the lowest validation loss.

For each CNN model, the training phase was performed five times with different training/validation splits, generating five instances with different adjusted weights. The same five training/validation splits were used for all models. Each instance was evaluated on the test subset and the following measures were obtained: accuracy (ACC), sensitivity (TPR), precision (PPV), and F1 score. The results are shown in Table~\ref{tag:CNNResults}. Each value is the average of the measures obtained on the five different instances of each model. The standard deviation is also presented. Results of the same evaluation applied to the training and validation subsets are available in \ref{sec:AppendixSubsets}.

\begin{table}

\resizebox{\textwidth}{!}{%

\begin{tabular}{r|cc|cc|cc|cc}

\toprule

\multicolumn{ 1}{c|}{{\bf Model}} & \multicolumn{ 2}{c|}{{\bf ACC}} & \multicolumn{ 2}{c|}{{\bf TPR}} & \multicolumn{ 2}{c|}{{\bf PPV}} & \multicolumn{ 2}{c}{{\bf F1}} \\

\multicolumn{ 1}{c|}{{\bf }} & {\bf Mean} & {\bf S.D.} & {\bf Mean} & {\bf S.D.} & {\bf Mean} & {\bf S.D.} & {\bf Mean} & {\bf S.D.} \\

\toprule

DenseNet169 & {\bf 0.9815} &     0.0056 & {\bf 0.9700} &     0.0138 &     0.9930 &     0.0075 & {\bf 0.9812} &     0.0058 \\

EfficientNetB2 &     0.9760 &     0.0049 &     0.9600 &     0.0141 &     0.9918 &     0.0051 &     0.9756 &     0.0052 \\

InceptionResNetV2 &     0.9755 &     0.0099 &     0.9590 &     0.0246 &     0.9919 &     0.0051 &     0.9749 &     0.0106 \\

InceptionV3 &     0.9750 &     0.0065 &     0.9520 &     0.0144 &     0.9979 &     0.0041 &     0.9744 &     0.0069 \\

 MobileNet &     0.9710 &     0.0060 &     0.9430 &     0.0136 &     0.9990 &     0.0021 &     0.9701 &     0.0064 \\

EfficientNetB0 &     0.9705 &     0.0051 &     0.9510 &     0.0086 &     0.9896 &     0.0033 &     0.9699 &     0.0053 \\

EfficientNetB3 &     0.9700 &     0.0163 &     0.9470 &     0.0337 &     0.9927 &     0.0051 &     0.9690 &     0.0177 \\

DenseNet201 &     0.9695 &     0.0176 &     0.9400 &     0.0342 &     0.9989 &     0.0022 &     0.9683 &     0.0186 \\

ResNet152V2 &     0.9695 &     0.0244 &     0.9420 &     0.0510 &     0.9970 &     0.0040 &     0.9679 &     0.0268 \\

 ResNet152 &     0.9660 &     0.0223 &     0.9370 &     0.0443 &     0.9947 &     0.0033 &     0.9644 &     0.0243 \\

DenseNet121 &     0.9630 &     0.0053 &     0.9270 &     0.0103 &     0.9989 &     0.0022 &     0.9616 &     0.0057 \\

  Xception &     0.9615 &     0.0077 &     0.9230 &     0.0154 & {\bf 1.0000} &     0.0000 &     0.9599 &     0.0083 \\

     VGG19 &     0.9580 &     0.0198 &     0.9170 &     0.0385 &     0.9989 &     0.0023 &     0.9558 &     0.0216 \\

EfficientNetB1 &     0.9570 &     0.0224 &     0.9240 &     0.0413 &     0.9892 &     0.0075 &     0.9551 &     0.0242 \\

  ResNet50 &     0.9545 &     0.0172 &     0.9090 &     0.0344 & {\bf 1.0000} &     0.0000 &     0.9520 &     0.0192 \\

     VGG16 &     0.9525 &     0.0123 &     0.9090 &     0.0282 &     0.9958 &     0.0052 &     0.9501 &     0.0138 \\

ResNet101V2 &     0.9530 &     0.0302 &     0.9100 &     0.0643 &     0.9959 &     0.0050 &     0.9497 &     0.0342 \\

MobileNetV2 &     0.9485 &     0.0172 &     0.9030 &     0.0359 &     0.9935 &     0.0019 &     0.9457 &     0.0190 \\

 ResNet101 &     0.9410 &     0.0170 &     0.8830 &     0.0333 &     0.9988 &     0.0023 &     0.9370 &     0.0190 \\

ResNet50V2 &     0.9280 &     0.0075 &     0.8590 &     0.0153 &     0.9966 &     0.0046 &     0.9226 &     0.0087 \\

NASNetMobile &     0.8530 &     0.0653 &     0.7090 &     0.1317 &     0.9960 &     0.0034 &     0.8212 &     0.0918 \\

\midrule

{\it Average} &     0.9569 &     0.0162 &     0.9178 &     0.0334 &     0.9957 &     0.0036 &     0.9536 &     0.0187 \\

\bottomrule

\end{tabular}

}

\caption{Comparison of $21$ different CNN models applied to the COVIDx8B dataset. Each model is executed five times. The highest values for each measure are highlighted in bold.}
\label{tag:CNNResults}
\end{table}

It is worth noticing that most related work only shows the results of a single execution on each tested CNN architecture. This may lead to wrong conclusions as there is always some expected variance on multiple executions of neural networks, which are stochastic by nature.

DenseNet169 achieved the highest accuracy ($98.15\%$), TPR ($97.00\%$), and F1 score ($98.12\%$) among all the tested models. The highest PPV ($100\%$) was achieved by Xception and ResNet50 models. EfficientNetB2 achieved the second-best accuracy, PPV, and F1 score.

Compared to other recent approaches applied to the same dataset, DenseNet169, EfficientNetB2, and InceptionResNetV2 achieved the best accuracy, TPR, and F1 score, as shown in Table~\ref{tab:CNNComparison}. It is worth noticing that EfficientNetB2 has less trainable parameters ($8.06$ million) than all the other architectures in this comparison, including the Covid-Net CXR-2 ($9.2$ million), which was specially tailored for the COVIDx8B dataset.

There are some common characteristics among the two best-performing CNN architectures. DenseNet and EfficientNet are newer approaches (2017 and 2019, respectively) than VGG (2015) and ResNet (2016). DenseNet and EfficientNet also focus on architecture efficiency to use less trainable parameters than the earlier approaches. In this case, the strategy used in these newer models was more suitable for these types of CXR images. Unfortunately, many related works compared fewer and/or earlier models only. Therefore, future studies should consider a wider variety of models to verify if this tendency confirms. In particular, from the related work section, only \citet{Nigam2021} and \citet{Monshi2021} explored EfficientNet, but they also reported good results with it, showing this is a promising architecture for CXR images.

\begin{table}

\resizebox{\textwidth}{!}{%

\begin{tabular}{r|c|c|c|c|l}

\toprule

{\bf Model}  &  {\bf ACC} &  {\bf TPR} &  {\bf PPV} &   {\bf F1} & {\bf Source} \\

\toprule

{\emph{DenseNet169}} & {\bf 0.9815} & {\bf 0.9700} &     0.9930 & {\bf 0.9812} & \emph{this paper} \\

{\emph{EfficientNetB2}}  &     0.9760 &     0.9600 &     0.9918 &     0.9756  & {\emph{this paper}} \\

{\emph{InceptionResNetV2}} &     0.9755 &     0.9590 &     0.9919 &     0.9749 & {\emph{this paper}} \\

{\emph{InceptionV3}} &     0.9750 &     0.9520 &     0.9979 &     0.9744 & {\emph{this paper}} \\

VGG16 (ImageNet)  &     0.9750 &     0.9500 & {\bf 1.0000} &     0.9744  & \citet{Dominik2021}  \\

 Covid-Net  &     0.9400 &     0.9350 & {\bf 1.0000} &     0.9664 & \citet{Pavlova2021} \\

DenseNet121 (ChestXray)  &     0.9650 &     0.9350 &     0.9947 &     0.9639  & \citet{Dominik2021} \\

ResNet50V2 (Bit-M)  &     0.9650 &     0.9300 & {\bf 1.0000} &     0.9637 &   \citet{Zhao2021} \\

Covid-Net CXR-2  &     0.9630 &     0.9550 &     0.9700 &     0.9624 & \citet{Pavlova2021} \\

VGG19 (ImageNet)  &     0.9625 &     0.9250 & {\bf 1.0000} &     0.9610 & \citet{Dominik2021}  \\

ResNet-50 (ImageNet)  &     0.9575 &     0.9200 &     0.9946 &     0.9558 & \citet{Dominik2021} \\

DenseNet121 (ImageNet)  &     0.9575 &     0.9150 & {\bf 1.0000} &     0.9556 & \citet{Dominik2021} \\

Xception (ImageNet)  &     0.9550 &     0.9100 & {\bf 1.0000} &     0.9529  & \citet{Dominik2021} \\

ResNet50V2 (Bit-S)  &     0.9480 &     0.8950 & {\bf 1.0000} &     0.9446  &   \citet{Zhao2021} \\

ResNet50V2 (Random)  &     0.9280 &     0.8550 & {\bf 1.0000} &     0.9218  &   \citet{Zhao2021} \\

  ResNet50  &     0.9050 &     0.8850 &     0.9220 &     0.9031  & \citet{Pavlova2021} \\

\bottomrule

\end{tabular}

}

\caption{Comparison of the best four models tested in this paper (in italic) with other recently proposed models applied to the COVIDx8B dataset. The highest values for each measure are highlighted in bold. The results obtained by other authors were compiled from the respective cited references. }

\label{tab:CNNComparison}

\end{table}

Table~\ref{tab:CNNComparisonRelated} compares results reported in individual papers described in Section~\ref{sec:RelatedWork}, where the authors are motivated to use a setup such their algorithm is the best performing, with the best result found in this paper for an individual CNN architecture, in which there is no motivation to implement optimizations to boost any particular architecture. Despite that, the best result from this paper is still in the top half of the best accuracy ranking. For each paper, the CNN architecture used and the dataset size are provided for reference.

\begin{table}

\begin{tabular}{llcc}

\toprule

{\bf Reference} & {\bf Architecture} & {\bf Dataset Size} & {\bf Accuracy} \\

\toprule

\citet{Alawad2021} &      VGG16 &      7,329 &    99.82\% \\

\citet{Karthik2021} &       CSDB &     15,265 &    99.80\% \\

\citet{Hira2021} & Se-ResNeXt-50 &      8,830 &    99.32\% \\

\citet{Jia2021} &  MobileNet &      7,592 &    99.30\% \\

\citet{Mostafiz2020} &   ResNet50 &      4,809 &    98.50\% \\

\citet{Narin2021} &   ResNet50 &      7.406 &    98.43\% \\

{\it this paper} & {\it DenseNet169} &     16,352 &    98.15\% \\

\citet{Chhikara2021} & InceptionV3 &     11,244 &    97.70\% \\

\citet{Chhikara2021} & InceptionV3 &     14,486 &    97.03\% \\

\citet{Monshi2021} & EfficientNetB0 &     15,496 &    95.82\% \\

\citet{Jia2021} &  MobileNet &     13.975 &    95,00\% \\

\citet{Ismael2021} &   ResNet50 &        380 &    94.70\% \\

\citet{Heidari2020} &      VGG16 &      8,474 &    94.50\% \\

\citet{Nigam2021} & EfficientNetB7 &     16,634 &    93.48\% \\

\citet{Abbas2021} &     DeTrac &        196 &    93.10\% \\

\citet{Chhikara2021} & InceptionV3 &      8,246 &    84.95\% \\

\bottomrule

\end{tabular}

\caption{Comparison of different CNN-based models applied to different COVID-19 datasets found in individual papers and the best result by an individual CNN model applied to the COVIDx8B dataset in this paper.}
\label{tab:CNNComparisonRelated}
\end{table}

\section{CNN Ensembles}
\label{sec:CNNEnsembles}

This section presents the computer simulations with ensembles of different CNN models and ensembles of multiple instances of the same model. All the ensembles experiments used the output of the last dense layer, just before the softmax activation function. Therefore, for each image, each model will output two continuous values, which can be interpreted as the probability of each class. Then, the output of the ensemble will be the average of its members' output. The same weights trained for the experiments in Section~\ref{sec:CNNComparison} were used for the experiments in this section.

In the first ensemble experiment, the two models that achieved the best individual F1 score (DenseNet169 and EfficientNetB2) were combined in the first ensemble configuration. The second ensemble configuration adds the third-best model (InceptionResNetV2). The third ensemble configuration adds the fourth-best model (InceptionV3) and so on, with up to seven models. Then, in the last ensemble configuration, all the $21$ models are combined. In this first experiment, only one instance of each model composes each ensemble, thus there are five ensembles for each configuration. Table~\ref{tab:Ensembles} shows the average and standard deviation of the measures obtained for each ensemble configuration.

The best accuracy, TPR, and F1 score were achieved when the three best models were combined (DenseNet169, EfficientNetB2, and InceptionResNetV2). All the ensembles, except for the one with the best two models, achieved a PPV of $100\%$. Except for the ensemble of all models, all the other ensembles achieved higher accuracy, TPR, and F1 scores than the best individual model.

\begin{table}

\resizebox{\textwidth}{!}{%

\begin{tabular}{r|cc|cc|cc|cc}

\toprule

\multicolumn{ 1}{c|}{{\bf Models}} & \multicolumn{ 2}{c|}{{\bf ACC}} & \multicolumn{ 2}{c|}{{\bf TPR}} & \multicolumn{ 2}{c|}{{\bf PPV}} & \multicolumn{ 2}{c}{{\bf F1}} \\

\multicolumn{ 1}{c|}{{\bf }} & {\bf Mean} & {\bf S.D.} & {\bf Mean} & {\bf S.D.} & {\bf Mean} & {\bf S.D.} & {\bf Mean} & {\bf S.D.} \\

\toprule

Top 2 models &     0.9855 &     0.0024 &     0.9730 &     0.0040 &     0.9980 &     0.0025 &     0.9853 &     0.0025 \\

Top 3 models & {\bf 0.9885} &   0.0034 &     0.9770 &     0.0068 & {\bf 1.0000} &     0.0000 &     0.9884 &     0.0035 \\

Top 4 models &     0.9870 &     0.0019 &     0.9740 &     0.0037 & {\bf 1.0000} &     0.0000 &     0.9868 &     0.0019 \\

Top 5 models &     0.9865 &     0.0020 &     0.9730 &     0.0040 & {\bf 1.0000} &     0.0000 &     0.9863 &     0.0021 \\

Top 6 models &     0.9880 &     0.0010 & {\bf 0.9760} &     0.0020 & {\bf 1.0000} &     0.0000 & {\bf 0.9879} &     0.0010 \\

Top 7 models &     0.9865 &     0.0025 &     0.9730 &     0.0051 & {\bf 1.0000} &     0.0000 &     0.9863 &     0.0026 \\

All models &     0.9775 &     0.0032 &     0.9550 &     0.0063 & {\bf 1.0000} &     0.0000 &     0.9770 &     0.0033 \\

\bottomrule

\end{tabular}

}

\caption{Ensembles of CNN models applied to the COVIDx8B dataset. Each ensemble configuration is executed five times with different instances of the models. The highest values for each measure are highlighted in bold.}

\label{tab:Ensembles}

\end{table}

For the second ensembles experiment, the five instances of each model are combined to form an ensemble. It is expected that five instances, even if they are from the same model, will improve the measures by alleviating the randomness effects of the training.

Table~\ref{tab:EnsembleSameModel} shows the measures obtained with these ensembles for each model. It also shows the gain obtained by the ensemble when compared to the average of the single instances. All the models had gained with the ensembles. The highest measures were obtained by DenseNet169, with an F1 score of $99.24\%$ and an accuracy of $99.25\%$. This is the same accuracy obtained by \citet{Dominik2021} using an ensemble of multiple models and an optimized threshold. To the best of my knowledge, this is the highest accuracy achieved in this dataset at the time this paper is being written.

\begin{table}

\resizebox{\textwidth}{!}{%

\begin{tabular}{r|cc|cc|cc|cc}

\toprule

{\bf Models} & \multicolumn{ 2}{c|}{{\bf ACC}} & \multicolumn{ 2}{c|}{{\bf TPR}} & \multicolumn{ 2}{c|}{{\bf PPV}} & \multicolumn{ 2}{c}{{\bf F1}} \\

           & {\bf Mean} & {\bf Gain} & {\bf Mean} & {\bf Gain} & {\bf Mean} & {\bf Gain} & {\bf Mean} & {\bf Gain} \\

\toprule

DenseNet169 & {\bf 0.9925} &     1.12\% & {\bf 0.9850} &     1.55\% & {\bf 1.0000} &  {\bf 0.70\%} & {\bf 0.9924} &     1.14\% \\

EfficientNetB2 &     0.9850 &     0.92\% &     0.9750 &     1.56\% &     0.9949 &     0.31\% &     0.9848 &     0.94\% \\

InceptionResNetV2 &     0.9875 &     1.23\% &     0.9750 &     1.67\% & {\bf 1.0000} &     0.82\% &     0.9873 &     1.27\% \\

InceptionV3 &     0.9800 &     0.51\% &     0.9600 &     0.84\% & {\bf 1.0000} &     0.21\% &     0.9796 &     0.53\% \\

 MobileNet &     0.9825 &     1.18\% &     0.9650 &     2.33\% & {\bf 1.0000} &     0.10\% &     0.9822 &     1.25\% \\

EfficientNetB0 &     0.9750 &     0.46\% &     0.9600 &     0.95\% &     0.9897 &     0.01\% &     0.9746 &     0.48\% \\

EfficientNetB3 &     0.9850 &     1.55\% &     0.9750 &     2.96\% &     0.9949 &     0.22\% &     0.9848 &     1.63\% \\

DenseNet201 &     0.9825 &     1.34\% &     0.9650 &     2.66\% & {\bf 1.0000} &     0.11\% &     0.9822 &     1.44\% \\

ResNet152V2 &     0.9900 &     2.11\% &     0.9800 &     4.03\% & {\bf 1.0000} &     0.30\% &     0.9899 &     2.27\% \\

 ResNet152 &     0.9800 &     1.45\% &     0.9650 &     2.99\% &     0.9948 &     0.01\% &     0.9797 &     1.59\% \\

DenseNet121 &     0.9725 &     0.99\% &     0.9450 &     1.94\% & {\bf 1.0000} &     0.11\% &     0.9717 &     1.05\% \\

  Xception &     0.9625 &     0.10\% &     0.9250 &     0.22\% & {\bf 1.0000} &     0.00\% &     0.9610 &     0.11\% \\

     VGG19 &     0.9700 &     1.25\% &     0.9400 &     2.51\% & {\bf 1.0000} &     0.11\% &     0.9691 &     1.39\% \\

EfficientNetB1 &     0.9725 &     1.62\% &     0.9500 &     2.81\% &     0.9948 &     0.57\% &     0.9719 &     1.76\% \\

  ResNet50 &     0.9650 &     1.10\% &     0.9300 &     2.31\% & {\bf 1.0000} &     0.00\% &     0.9637 &     1.23\% \\

     VGG16 &     0.9550 &     0.26\% &     0.9100 &     0.11\% & {\bf 1.0000} &     0.42\% &     0.9529 &     0.29\% \\

ResNet101V2 &     0.9650 &     1.26\% &     0.9300 &     2.20\% & {\bf 1.0000} &     0.41\% &     0.9637 &     1.47\% \\

MobileNetV2 &     0.9650 &     1.74\% &     0.9350 &     3.54\% &     0.9947 &     0.12\% &     0.9639 &     1.92\% \\

 ResNet101 &     0.9575 &     1.75\% &     0.9150 &     3.62\% & {\bf 1.0000} &     0.12\% &     0.9556 &     1.99\% \\

ResNet50V2 &     0.9350 &     0.75\% &     0.8700 &     1.28\% & {\bf 1.0000} &     0.34\% &     0.9305 &     0.86\% \\

NASNetMobile &     0.8750 &    {\bf 2.58\%} &     0.7500 &   {\bf 5.78\%} & {\bf 1.0000} &     0.40\% &     0.8571 &   {\bf  4.37\%} \\

\midrule

{\it Average} &     0.9683 &     1.20\% &     0.9383 &     2.28\% &     0.9983 &     0.26\% &     0.9666 &     1.38\% \\

\bottomrule

\end{tabular}

}

\caption{Ensembles of CNN models applied to the COVIDx8B dataset. Each ensemble is composed of five instances of the same model, with different training/validation splits. The highest values for each measure and the highest gains in comparison to single instances of each model are highlighted in bold.}

\label{tab:EnsembleSameModel}

\end{table}

For the third and last ensembles experiment, the first experiment is repeated, but now using all the five instances of each model in the ensemble. Table~\ref{tab:EnsembleAllInstances} shows the measures obtained with each ensemble and the gain obtained by these ensembles when compared to the ensembles which used only a single instance of each model. In this case, there were only small differences and some of them were negative. Therefore, the best ensemble overall is still the one with multiple instances of DenseNet169.

\begin{table}

\resizebox{\textwidth}{!}{%

\begin{tabular}{r|cc|cc|cc|cc}

\toprule

\multicolumn{ 1}{c|}{{\bf Models}} & \multicolumn{ 2}{c|}{{\bf ACC}} & \multicolumn{ 2}{c|}{{\bf TPR}} & \multicolumn{ 2}{c|}{{\bf PPV}} & \multicolumn{ 2}{c}{{\bf F1}} \\

\multicolumn{ 1}{c|}{{\bf }} & {\bf Mean} & {\bf Gain} & {\bf Mean} & {\bf Gain} & {\bf Mean} & {\bf Gain} & {\bf Mean} & {\bf Gain} \\

\toprule

Top 2 models &     0.9850 &    -0.05\% &     0.9700 &    -0.31\% & {\bf 1.0000} & {\bf 0.20\%} &     0.9848 &    -0.05\% \\

Top 3 models & {\bf 0.9875} &     -0.10\% & {\bf 0.9750} &    -0.20\% & {\bf 1.0000} &     0.00\% & {\bf 0.9873} &    -0.11\% \\

Top 4 models & {\bf 0.9875} &     0.05\% & {\bf 0.9750} &     0.10\% & {\bf 1.0000} &     0.00\% & {\bf 0.9873} &     0.05\% \\

Top 5 models & {\bf 0.9875} & {\bf 0.10\%} & {\bf 0.9750} & {\bf 0.21\%} & {\bf 1.0000} &     0.00\% & {\bf 0.9873} & {\bf 0.10\%} \\

Top 6 models & {\bf 0.9875} &    -0.05\% & {\bf 0.9750} &    -0.10\% & {\bf 1.0000} &     0.00\% & {\bf 0.9873} &    -0.06\% \\

Top 7 models & {\bf 0.9875} & {\bf 0.10\%} & {\bf 0.9750} & {\bf 0.21\%} & {\bf 1.0000} &     0.00\% & {\bf 0.9873} & {\bf 0.10\%} \\

All models &     0.9775 &     0.00\% &     0.9550 &     0.00\% & {\bf  1.0000} &     0.00\% &     0.9770 &     0.00\% \\

\midrule

{\it Average} &     0.9857 &     0.01\% &     0.9714 &    -0.01\% &     1.0000 &     0.03\% &     0.9855 &     0.00\% \\

\bottomrule

\end{tabular}

}

\caption{Ensembles of CNN models applied to the COVIDx8B dataset. Each ensemble configuration has five instances of each participant model. The highest values for each measure and the highest gains in comparison with the ensembles of single instances for each model are highlighted in bold.}

\label{tab:EnsembleAllInstances}

\end{table}

\section{Conclusions}
\label{sec:Conclusions}

In this paper, $21$ different CNN architectures are applied to the detection of COVID-19 on CXR images. The comparison was performed using the COVIDx8B, a large and heterogeneous COVID-19 CXR images dataset, which is composed of six open-source CXR datasets. The training was repeated five times for each model, with different training and validation splits to get more reliable results, while most related works tested fewer models and performed only a single execution for each one.

CNN ensembles were also explored in this work, combining both different models and multiple instances of the same model. DenseNet169 achieved the best results regarding the accuracy and the F1 score, both as a single instance and with an ensemble of five instances. The classification accuracies were $98.15\%$ and $99.25\%$ for the single instance and the ensemble, respectively, while the F1 scores were $98.12\%$ and $99.24\%$, also respectively. These results are better than those achieved in recent works where the same dataset was used.

The simulations performed for this paper add more evidence of the efficacy of CNNs in the detection of COVID-19 on CXR images, which is very important to assist in quick diagnostics and to avoid the spread of the disease. Moreover, these experiments may also guide future research as they tested a large amount of CNN architectures and identified which of them produces the best results for this particular task.


\bibliographystyle{elsarticle-harv}
\bibliography{covid-cnn}

\begin{thebibliography}{46}
\expandafter\ifx\csname natexlab\endcsname\relax\def\natexlab#1{#1}\fi
\providecommand{\url}[1]{\texttt{#1}}
\providecommand{\href}[2]{#2}
\providecommand{\path}[1]{#1}
\providecommand{\DOIprefix}{doi:}
\providecommand{\ArXivprefix}{arXiv:}
\providecommand{\URLprefix}{URL: }
\providecommand{\Pubmedprefix}{pmid:}
\providecommand{\doi}[1]{\href{http://dx.doi.org/#1}{\path{#1}}}
\providecommand{\Pubmed}[1]{\href{pmid:#1}{\path{#1}}}
\providecommand{\bibinfo}[2]{#2}
\ifx\xfnm\relax \def\xfnm[#1]{\unskip,\space#1}\fi
\bibitem[{Abbas et~al.(2021)Abbas, Abdelsamea and Gaber}]{Abbas2021}
\bibinfo{author}{Abbas, A.}, \bibinfo{author}{Abdelsamea, M.M.},
  \bibinfo{author}{Gaber, M.M.}, \bibinfo{year}{2021}.
\newblock \bibinfo{title}{Classification of covid-19 in chest x-ray images
  using detrac deep convolutional neural network}.
\newblock \bibinfo{journal}{Applied Intelligence} \bibinfo{volume}{51},
  \bibinfo{pages}{854--864}.
\bibitem[{Alawad et~al.(2021)Alawad, Alburaidi, Alzahrani and
  Alflaj}]{Alawad2021}
\bibinfo{author}{Alawad, W.}, \bibinfo{author}{Alburaidi, B.},
  \bibinfo{author}{Alzahrani, A.}, \bibinfo{author}{Alflaj, F.},
  \bibinfo{year}{2021}.
\newblock \bibinfo{title}{A comparative study of stand-alone and hybrid cnn
  models for covid-19 detection}.
\newblock \bibinfo{journal}{International Journal of Advanced Computer Science
  and Applications} \bibinfo{volume}{12}.
\newblock \URLprefix \url{http://dx.doi.org/10.14569/IJACSA.2021.01206102},
  \DOIprefix\doi{10.14569/IJACSA.2021.01206102}.
\bibitem[{Arevalo-Rodriguez et~al.(2020)Arevalo-Rodriguez, Buitrago-Garcia,
  Simancas-Racines, Zambrano-Achig, Del~Campo, Ciapponi, Sued, Martinez-Garcia,
  Rutjes, Low et~al.}]{Arevalo2020}
\bibinfo{author}{Arevalo-Rodriguez, I.}, \bibinfo{author}{Buitrago-Garcia, D.},
  \bibinfo{author}{Simancas-Racines, D.}, \bibinfo{author}{Zambrano-Achig, P.},
  \bibinfo{author}{Del~Campo, R.}, \bibinfo{author}{Ciapponi, A.},
  \bibinfo{author}{Sued, O.}, \bibinfo{author}{Martinez-Garcia, L.},
  \bibinfo{author}{Rutjes, A.W.}, \bibinfo{author}{Low, N.}, et~al.,
  \bibinfo{year}{2020}.
\newblock \bibinfo{title}{False-negative results of initial rt-pcr assays for
  covid-19: a systematic review}.
\newblock \bibinfo{journal}{PloS one} \bibinfo{volume}{15},
  \bibinfo{pages}{e0242958}.
\bibitem[{Chhikara et~al.(2021)Chhikara, Gupta, Singh and
  Bhatia}]{Chhikara2021}
\bibinfo{author}{Chhikara, P.}, \bibinfo{author}{Gupta, P.},
  \bibinfo{author}{Singh, P.}, \bibinfo{author}{Bhatia, T.},
  \bibinfo{year}{2021}.
\newblock \bibinfo{title}{A deep transfer learning based model for automatic
  detection of covid-19 from chest x-rays}.
\newblock \bibinfo{journal}{Turkish Journal of Electrical Engineering \&
  Computer Sciences} \bibinfo{volume}{29}, \bibinfo{pages}{2663--2679}.
\bibitem[{{Chollet}(2017)}]{Chollet2017}
\bibinfo{author}{{Chollet}, F.}, \bibinfo{year}{2017}.
\newblock \bibinfo{title}{Xception: Deep learning with depthwise separable
  convolutions}, in: \bibinfo{booktitle}{2017 IEEE Conference on Computer
  Vision and Pattern Recognition (CVPR)}, pp. \bibinfo{pages}{1800--1807}.
\newblock \DOIprefix\doi{10.1109/CVPR.2017.195}.
\bibitem[{Chowdhury et~al.(2020)Chowdhury, Rahman, Khandakar, Mazhar, Kadir,
  Mahbub, Islam, Khan, Iqbal, Emadi, Reaz and Islam}]{Chowdhury2020}
\bibinfo{author}{Chowdhury, M.E.H.}, \bibinfo{author}{Rahman, T.},
  \bibinfo{author}{Khandakar, A.}, \bibinfo{author}{Mazhar, R.},
  \bibinfo{author}{Kadir, M.A.}, \bibinfo{author}{Mahbub, Z.B.},
  \bibinfo{author}{Islam, K.R.}, \bibinfo{author}{Khan, M.S.},
  \bibinfo{author}{Iqbal, A.}, \bibinfo{author}{Emadi, N.A.},
  \bibinfo{author}{Reaz, M.B.I.}, \bibinfo{author}{Islam, M.T.},
  \bibinfo{year}{2020}.
\newblock \bibinfo{title}{Can ai help in screening viral and covid-19
  pneumonia?}
\newblock \bibinfo{journal}{IEEE Access} \bibinfo{volume}{8},
  \bibinfo{pages}{132665--132676}.
\newblock \DOIprefix\doi{10.1109/ACCESS.2020.3010287}.
\bibitem[{Chung(2020a)}]{Chung2020b}
\bibinfo{author}{Chung, A.}, \bibinfo{year}{2020}a.
\newblock \bibinfo{title}{Actualmed covid-19 chest x-ray dataset initiative}.
\newblock
  \bibinfo{howpublished}{\url{https://github.com/agchung/Actualmed-COVID-chestxray-dataset}}.
\bibitem[{Chung(2020b)}]{Chung2020a}
\bibinfo{author}{Chung, A.}, \bibinfo{year}{2020}b.
\newblock \bibinfo{title}{Figure1-covid-chestxray-dataset}.
\newblock
  \bibinfo{howpublished}{\url{https://github.com/agchung/Figure1-COVID-chestxray-dataset}}.
\bibitem[{Cohen et~al.(2020)Cohen, Morrison, Dao, Roth, Duong and
  Ghassemi}]{Cohen2020}
\bibinfo{author}{Cohen, J.P.}, \bibinfo{author}{Morrison, P.},
  \bibinfo{author}{Dao, L.}, \bibinfo{author}{Roth, K.},
  \bibinfo{author}{Duong, T.Q.}, \bibinfo{author}{Ghassemi, M.},
  \bibinfo{year}{2020}.
\newblock \bibinfo{title}{Covid-19 image data collection: Prospective
  predictions are the future}.
\newblock \bibinfo{journal}{arXiv 2006.11988} \URLprefix
  \url{https://github.com/ieee8023/covid-chestxray-dataset}.
\bibitem[{Dominik(2021)}]{Dominik2021}
\bibinfo{author}{Dominik, C.}, \bibinfo{year}{2021}.
\newblock \bibinfo{title}{Detection of COVID-19 in X-ray images using Neural
  Networks}.
\newblock \bibinfo{type}{Bachelor's thesis}. Czech Technical University in
  Prague, Faculty of Information Technology.
\bibitem[{Feng et~al.(2020)Feng, Liu, Lv and Zhong}]{Feng2020}
\bibinfo{author}{Feng, H.}, \bibinfo{author}{Liu, Y.}, \bibinfo{author}{Lv,
  M.}, \bibinfo{author}{Zhong, J.}, \bibinfo{year}{2020}.
\newblock \bibinfo{title}{A case report of covid-19 with false negative rt-pcr
  test: necessity of chest ct}.
\newblock \bibinfo{journal}{Japanese journal of radiology}
  \bibinfo{volume}{38}, \bibinfo{pages}{409--410}.
\bibitem[{Goodfellow et~al.(2016)Goodfellow, Bengio and
  Courville}]{Goodfellow2016}
\bibinfo{author}{Goodfellow, I.}, \bibinfo{author}{Bengio, Y.},
  \bibinfo{author}{Courville, A.}, \bibinfo{year}{2016}.
\newblock \bibinfo{title}{Deep learning}.
\newblock \bibinfo{publisher}{MIT press}.
\bibitem[{He et~al.(2016a)He, Zhang, Ren and Sun}]{He2016}
\bibinfo{author}{He, K.}, \bibinfo{author}{Zhang, X.}, \bibinfo{author}{Ren,
  S.}, \bibinfo{author}{Sun, J.}, \bibinfo{year}{2016}a.
\newblock \bibinfo{title}{Deep residual learning for image recognition}, in:
  \bibinfo{booktitle}{The IEEE Conference on Computer Vision and Pattern
  Recognition (CVPR)}, pp. \bibinfo{pages}{770--778}.
\bibitem[{He et~al.(2016b)He, Zhang, Ren and Sun}]{He2016ResNetV2}
\bibinfo{author}{He, K.}, \bibinfo{author}{Zhang, X.}, \bibinfo{author}{Ren,
  S.}, \bibinfo{author}{Sun, J.}, \bibinfo{year}{2016}b.
\newblock \bibinfo{title}{Identity mappings in deep residual networks}, in:
  \bibinfo{editor}{Leibe, B.}, \bibinfo{editor}{Matas, J.},
  \bibinfo{editor}{Sebe, N.}, \bibinfo{editor}{Welling, M.} (Eds.),
  \bibinfo{booktitle}{Computer Vision -- ECCV 2016},
  \bibinfo{publisher}{Springer International Publishing},
  \bibinfo{address}{Cham}. pp. \bibinfo{pages}{630--645}.
\bibitem[{Heidari et~al.(2020)Heidari, Mirniaharikandehei, Khuzani, Danala, Qiu
  and Zheng}]{Heidari2020}
\bibinfo{author}{Heidari, M.}, \bibinfo{author}{Mirniaharikandehei, S.},
  \bibinfo{author}{Khuzani, A.Z.}, \bibinfo{author}{Danala, G.},
  \bibinfo{author}{Qiu, Y.}, \bibinfo{author}{Zheng, B.}, \bibinfo{year}{2020}.
\newblock \bibinfo{title}{Improving the performance of cnn to predict the
  likelihood of covid-19 using chest x-ray images with preprocessing
  algorithms}.
\newblock \bibinfo{journal}{International Journal of Medical Informatics}
  \bibinfo{volume}{144}, \bibinfo{pages}{104284}.
\newblock \URLprefix
  \url{https://www.sciencedirect.com/science/article/pii/S138650562030959X},
  \DOIprefix\doi{https://doi.org/10.1016/j.ijmedinf.2020.104284}.
\bibitem[{Hira et~al.(2021)Hira, Bai and Hira}]{Hira2021}
\bibinfo{author}{Hira, S.}, \bibinfo{author}{Bai, A.}, \bibinfo{author}{Hira,
  S.}, \bibinfo{year}{2021}.
\newblock \bibinfo{title}{An automatic approach based on cnn architecture to
  detect covid-19 disease from chest x-ray images}.
\newblock \bibinfo{journal}{Applied Intelligence} \bibinfo{volume}{51},
  \bibinfo{pages}{2864--2889}.
\bibitem[{Howard et~al.(2017)Howard, Zhu, Chen, Kalenichenko, Wang, Weyand,
  Andreetto and Adam}]{Howard2017}
\bibinfo{author}{Howard, A.G.}, \bibinfo{author}{Zhu, M.},
  \bibinfo{author}{Chen, B.}, \bibinfo{author}{Kalenichenko, D.},
  \bibinfo{author}{Wang, W.}, \bibinfo{author}{Weyand, T.},
  \bibinfo{author}{Andreetto, M.}, \bibinfo{author}{Adam, H.},
  \bibinfo{year}{2017}.
\newblock \bibinfo{title}{Mobilenets: Efficient convolutional neural networks
  for mobile vision applications}.
\newblock \bibinfo{journal}{arXiv preprint arXiv:1704.04861} .
\bibitem[{Huang et~al.(2017)Huang, Liu, van~der Maaten and
  Weinberger}]{Huang2017}
\bibinfo{author}{Huang, G.}, \bibinfo{author}{Liu, Z.},
  \bibinfo{author}{van~der Maaten, L.}, \bibinfo{author}{Weinberger, K.Q.},
  \bibinfo{year}{2017}.
\newblock \bibinfo{title}{Densely connected convolutional networks}, in:
  \bibinfo{booktitle}{Proceedings of the IEEE Conference on Computer Vision and
  Pattern Recognition (CVPR)}.
\bibitem[{Ismael and Şengür(2021)}]{Ismael2021}
\bibinfo{author}{Ismael, A.M.}, \bibinfo{author}{Şengür, A.},
  \bibinfo{year}{2021}.
\newblock \bibinfo{title}{Deep learning approaches for covid-19 detection based
  on chest x-ray images}.
\newblock \bibinfo{journal}{Expert Systems with Applications}
  \bibinfo{volume}{164}, \bibinfo{pages}{114054}.
\newblock \URLprefix
  \url{https://www.sciencedirect.com/science/article/pii/S0957417420308198},
  \DOIprefix\doi{https://doi.org/10.1016/j.eswa.2020.114054}.
\bibitem[{Jia et~al.(2021)Jia, Lam and Xu}]{Jia2021}
\bibinfo{author}{Jia, G.}, \bibinfo{author}{Lam, H.K.}, \bibinfo{author}{Xu,
  Y.}, \bibinfo{year}{2021}.
\newblock \bibinfo{title}{Classification of covid-19 chest x-ray and ct images
  using a type of dynamic cnn modification method}.
\newblock \bibinfo{journal}{Computers in Biology and Medicine}
  \bibinfo{volume}{134}, \bibinfo{pages}{104425}.
\newblock \URLprefix
  \url{https://www.sciencedirect.com/science/article/pii/S0010482521002195},
  \DOIprefix\doi{https://doi.org/10.1016/j.compbiomed.2021.104425}.
\bibitem[{Karthik et~al.(2021)Karthik, Menaka and M.}]{Karthik2021}
\bibinfo{author}{Karthik, R.}, \bibinfo{author}{Menaka, R.},
  \bibinfo{author}{M., H.}, \bibinfo{year}{2021}.
\newblock \bibinfo{title}{Learning distinctive filters for covid-19 detection
  from chest x-ray using shuffled residual cnn}.
\newblock \bibinfo{journal}{Applied Soft Computing} \bibinfo{volume}{99},
  \bibinfo{pages}{106744}.
\newblock \URLprefix
  \url{https://www.sciencedirect.com/science/article/pii/S1568494620306827},
  \DOIprefix\doi{https://doi.org/10.1016/j.asoc.2020.106744}.
\bibitem[{Khan et~al.(2021)Khan, Mehran, Haq, Ullah, Naqvi, Ihsan and
  Abbass}]{Khan2021}
\bibinfo{author}{Khan, M.}, \bibinfo{author}{Mehran, M.T.},
  \bibinfo{author}{Haq, Z.U.}, \bibinfo{author}{Ullah, Z.},
  \bibinfo{author}{Naqvi, S.R.}, \bibinfo{author}{Ihsan, M.},
  \bibinfo{author}{Abbass, H.}, \bibinfo{year}{2021}.
\newblock \bibinfo{title}{Applications of artificial intelligence in covid-19
  pandemic: A comprehensive review}.
\newblock \bibinfo{journal}{Expert Systems with Applications}
  \bibinfo{volume}{185}, \bibinfo{pages}{115695}.
\newblock \URLprefix
  \url{https://www.sciencedirect.com/science/article/pii/S0957417421010794},
  \DOIprefix\doi{https://doi.org/10.1016/j.eswa.2021.115695}.
\bibitem[{Kingma and Ba(2014)}]{Kingma2014}
\bibinfo{author}{Kingma, D.P.}, \bibinfo{author}{Ba, J.}, \bibinfo{year}{2014}.
\newblock \bibinfo{title}{Adam: A method for stochastic optimization}.
\newblock \bibinfo{journal}{arXiv preprint arXiv:1412.6980} .
\bibitem[{Krizhevsky et~al.(2012)Krizhevsky, Sutskever and
  Hinton}]{Krizhevsky2012}
\bibinfo{author}{Krizhevsky, A.}, \bibinfo{author}{Sutskever, I.},
  \bibinfo{author}{Hinton, G.E.}, \bibinfo{year}{2012}.
\newblock \bibinfo{title}{Imagenet classification with deep convolutional
  neural networks}, in: \bibinfo{editor}{Pereira, F.}, \bibinfo{editor}{Burges,
  C.J.C.}, \bibinfo{editor}{Bottou, L.}, \bibinfo{editor}{Weinberger, K.Q.}
  (Eds.), \bibinfo{booktitle}{Advances in Neural Information Processing Systems
  25}. \bibinfo{publisher}{Curran Associates, Inc.}, pp.
  \bibinfo{pages}{1097--1105}.
\newblock \URLprefix
  \url{http://papers.nips.cc/paper/4824-imagenet-classification-with-deep-convolutional-neural-networks.pdf}.
\bibitem[{LeCun et~al.(2015)LeCun, Bengio and Hinton}]{Lecun2015}
\bibinfo{author}{LeCun, Y.}, \bibinfo{author}{Bengio, Y.},
  \bibinfo{author}{Hinton, G.}, \bibinfo{year}{2015}.
\newblock \bibinfo{title}{Deep learning}.
\newblock \bibinfo{journal}{nature} \bibinfo{volume}{521},
  \bibinfo{pages}{436}.
\bibitem[{Long et~al.(2020)Long, Gombar, Hogan, Greninger, O’Reilly-Shah,
  Bryson-Cahn, Stevens, Rustagi, Jerome, Kong, Zehnder, Shah, Weiss, Pinsky and
  Sunshine}]{Long2020}
\bibinfo{author}{Long, D.R.}, \bibinfo{author}{Gombar, S.},
  \bibinfo{author}{Hogan, C.A.}, \bibinfo{author}{Greninger, A.L.},
  \bibinfo{author}{O’Reilly-Shah, V.}, \bibinfo{author}{Bryson-Cahn, C.},
  \bibinfo{author}{Stevens, B.}, \bibinfo{author}{Rustagi, A.},
  \bibinfo{author}{Jerome, K.R.}, \bibinfo{author}{Kong, C.S.},
  \bibinfo{author}{Zehnder, J.}, \bibinfo{author}{Shah, N.H.},
  \bibinfo{author}{Weiss, N.S.}, \bibinfo{author}{Pinsky, B.A.},
  \bibinfo{author}{Sunshine, J.E.}, \bibinfo{year}{2020}.
\newblock \bibinfo{title}{{Occurrence and Timing of Subsequent Severe Acute
  Respiratory Syndrome Coronavirus 2 Reverse-transcription Polymerase Chain
  Reaction Positivity Among Initially Negative Patients}}.
\newblock \bibinfo{journal}{Clinical Infectious Diseases} \bibinfo{volume}{72},
  \bibinfo{pages}{323--326}.
\newblock \URLprefix \url{https://doi.org/10.1093/cid/ciaa722},
  \DOIprefix\doi{10.1093/cid/ciaa722},
  \href{http://arxiv.org/abs/https://academic.oup.com/cid/article-pdf/72/2/323/36115745/ciaa722.pdf}{{\tt
  arXiv:https://academic.oup.com/cid/article-pdf/72/2/323/36115745/ciaa722.pdf}}.
\bibitem[{Mohammad~Shorfuzzaman(2020)}]{Shorfuzzaman2020}
\bibinfo{author}{Mohammad~Shorfuzzaman, M.M.}, \bibinfo{year}{2020}.
\newblock \bibinfo{title}{On the detection of covid-19 from chest x-ray images
  using cnn-based transfer learning}.
\newblock \bibinfo{journal}{Computers, Materials \& Continua}
  \bibinfo{volume}{64}, \bibinfo{pages}{1359--1381}.
\newblock \URLprefix \url{http://www.techscience.com/cmc/v64n3/39434},
  \DOIprefix\doi{10.32604/cmc.2020.011326}.
\bibitem[{Monshi et~al.(2021)Monshi, Poon, Chung and Monshi}]{Monshi2021}
\bibinfo{author}{Monshi, M.M.A.}, \bibinfo{author}{Poon, J.},
  \bibinfo{author}{Chung, V.}, \bibinfo{author}{Monshi, F.M.},
  \bibinfo{year}{2021}.
\newblock \bibinfo{title}{Covidxraynet: Optimizing data augmentation and cnn
  hyperparameters for improved covid-19 detection from cxr}.
\newblock \bibinfo{journal}{Computers in Biology and Medicine}
  \bibinfo{volume}{133}, \bibinfo{pages}{104375}.
\newblock \URLprefix
  \url{https://www.sciencedirect.com/science/article/pii/S0010482521001694},
  \DOIprefix\doi{https://doi.org/10.1016/j.compbiomed.2021.104375}.
\bibitem[{Mostafiz et~al.(2020)Mostafiz, Uddin, Reza, Rahman
  et~al.}]{Mostafiz2020}
\bibinfo{author}{Mostafiz, R.}, \bibinfo{author}{Uddin, M.S.},
  \bibinfo{author}{Reza, M.M.}, \bibinfo{author}{Rahman, M.M.}, et~al.,
  \bibinfo{year}{2020}.
\newblock \bibinfo{title}{Covid-19 detection in chest x-ray through random
  forest classifier using a hybridization of deep cnn and dwt optimized
  features}.
\newblock \bibinfo{journal}{Journal of King Saud University-Computer and
  Information Sciences} .
\bibitem[{Narin et~al.(2021)Narin, Kaya and Pamuk}]{Narin2021}
\bibinfo{author}{Narin, A.}, \bibinfo{author}{Kaya, C.},
  \bibinfo{author}{Pamuk, Z.}, \bibinfo{year}{2021}.
\newblock \bibinfo{title}{Automatic detection of coronavirus disease (covid-19)
  using x-ray images and deep convolutional neural networks}.
\newblock \bibinfo{journal}{Pattern Analysis and Applications} ,
  \bibinfo{pages}{1--14}.
\bibitem[{Nigam et~al.(2021)Nigam, Nigam, Jain, Dodia, Arora and
  Annappa}]{Nigam2021}
\bibinfo{author}{Nigam, B.}, \bibinfo{author}{Nigam, A.},
  \bibinfo{author}{Jain, R.}, \bibinfo{author}{Dodia, S.},
  \bibinfo{author}{Arora, N.}, \bibinfo{author}{Annappa, B.},
  \bibinfo{year}{2021}.
\newblock \bibinfo{title}{Covid-19: Automatic detection from x-ray images by
  utilizing deep learning methods}.
\newblock \bibinfo{journal}{Expert Systems with Applications}
  \bibinfo{volume}{176}, \bibinfo{pages}{114883}.
\newblock \URLprefix
  \url{https://www.sciencedirect.com/science/article/pii/S0957417421003249},
  \DOIprefix\doi{https://doi.org/10.1016/j.eswa.2021.114883}.
\bibitem[{Oquab et~al.(2014)Oquab, Bottou, Laptev and Sivic}]{Oquab2014}
\bibinfo{author}{Oquab, M.}, \bibinfo{author}{Bottou, L.},
  \bibinfo{author}{Laptev, I.}, \bibinfo{author}{Sivic, J.},
  \bibinfo{year}{2014}.
\newblock \bibinfo{title}{Learning and transferring mid-level image
  representations using convolutional neural networks}, in:
  \bibinfo{booktitle}{The IEEE Conference on Computer Vision and Pattern
  Recognition (CVPR)}.
\bibitem[{Pavlova et~al.(2021)Pavlova, Terhljan, Chung, Zhao, Surana,
  Aboutalebi, Gunraj, Sabri, Alaref and Wong}]{Pavlova2021}
\bibinfo{author}{Pavlova, M.}, \bibinfo{author}{Terhljan, N.},
  \bibinfo{author}{Chung, A.G.}, \bibinfo{author}{Zhao, A.},
  \bibinfo{author}{Surana, S.}, \bibinfo{author}{Aboutalebi, H.},
  \bibinfo{author}{Gunraj, H.}, \bibinfo{author}{Sabri, A.},
  \bibinfo{author}{Alaref, A.}, \bibinfo{author}{Wong, A.},
  \bibinfo{year}{2021}.
\newblock \bibinfo{title}{Covid-net cxr-2: An enhanced deep convolutional
  neural network design for detection of covid-19 cases from chest x-ray
  images}.
\newblock \href{http://arxiv.org/abs/2105.06640}{{\tt arXiv:2105.06640}}.
\bibitem[{Rahman et~al.(2021)Rahman, Khandakar, Qiblawey, Tahir, Kiranyaz,
  {Abul Kashem}, Islam, {Al Maadeed}, Zughaier, Khan and
  Chowdhury}]{Rahman2021}
\bibinfo{author}{Rahman, T.}, \bibinfo{author}{Khandakar, A.},
  \bibinfo{author}{Qiblawey, Y.}, \bibinfo{author}{Tahir, A.},
  \bibinfo{author}{Kiranyaz, S.}, \bibinfo{author}{{Abul Kashem}, S.B.},
  \bibinfo{author}{Islam, M.T.}, \bibinfo{author}{{Al Maadeed}, S.},
  \bibinfo{author}{Zughaier, S.M.}, \bibinfo{author}{Khan, M.S.},
  \bibinfo{author}{Chowdhury, M.E.}, \bibinfo{year}{2021}.
\newblock \bibinfo{title}{Exploring the effect of image enhancement techniques
  on covid-19 detection using chest x-ray images}.
\newblock \bibinfo{journal}{Computers in Biology and Medicine}
  \bibinfo{volume}{132}, \bibinfo{pages}{104319}.
\newblock \URLprefix
  \url{https://www.sciencedirect.com/science/article/pii/S001048252100113X},
  \DOIprefix\doi{https://doi.org/10.1016/j.compbiomed.2021.104319}.
\bibitem[{Russakovsky et~al.(2015)Russakovsky, Deng, Su, Krause, Satheesh, Ma,
  Huang, Karpathy, Khosla, Bernstein, Berg and Fei-Fei}]{ILSVRC15}
\bibinfo{author}{Russakovsky, O.}, \bibinfo{author}{Deng, J.},
  \bibinfo{author}{Su, H.}, \bibinfo{author}{Krause, J.},
  \bibinfo{author}{Satheesh, S.}, \bibinfo{author}{Ma, S.},
  \bibinfo{author}{Huang, Z.}, \bibinfo{author}{Karpathy, A.},
  \bibinfo{author}{Khosla, A.}, \bibinfo{author}{Bernstein, M.},
  \bibinfo{author}{Berg, A.C.}, \bibinfo{author}{Fei-Fei, L.},
  \bibinfo{year}{2015}.
\newblock \bibinfo{title}{{ImageNet Large Scale Visual Recognition Challenge}}.
\newblock \bibinfo{journal}{International Journal of Computer Vision (IJCV)}
  \bibinfo{volume}{115}, \bibinfo{pages}{211--252}.
\newblock \DOIprefix\doi{10.1007/s11263-015-0816-y}.
\bibitem[{Sandler et~al.(2018)Sandler, Howard, Zhu, Zhmoginov and
  Chen}]{Sandler2018}
\bibinfo{author}{Sandler, M.}, \bibinfo{author}{Howard, A.},
  \bibinfo{author}{Zhu, M.}, \bibinfo{author}{Zhmoginov, A.},
  \bibinfo{author}{Chen, L.C.}, \bibinfo{year}{2018}.
\newblock \bibinfo{title}{Mobilenetv2: Inverted residuals and linear
  bottlenecks}, in: \bibinfo{booktitle}{The IEEE Conference on Computer Vision
  and Pattern Recognition (CVPR)}, pp. \bibinfo{pages}{4510--4520}.
\bibitem[{Schmidhuber(2015)}]{Schmidhuber2015}
\bibinfo{author}{Schmidhuber, J.}, \bibinfo{year}{2015}.
\newblock \bibinfo{title}{Deep learning in neural networks: An overview}.
\newblock \bibinfo{journal}{Neural networks} \bibinfo{volume}{61},
  \bibinfo{pages}{85--117}.
\bibitem[{Simonyan and Zisserman(2015)}]{Simonyan2015}
\bibinfo{author}{Simonyan, K.}, \bibinfo{author}{Zisserman, A.},
  \bibinfo{year}{2015}.
\newblock \bibinfo{title}{Very deep convolutional networks for large-scale
  image recognition}, \bibinfo{publisher}{Computational and Biological Learning
  Society}. pp. \bibinfo{pages}{1--14}.
\bibitem[{Szegedy et~al.(2017)Szegedy, Ioffe, Vanhoucke and
  Alemi}]{Szegedy2017}
\bibinfo{author}{Szegedy, C.}, \bibinfo{author}{Ioffe, S.},
  \bibinfo{author}{Vanhoucke, V.}, \bibinfo{author}{Alemi, A.A.},
  \bibinfo{year}{2017}.
\newblock \bibinfo{title}{Inception-v4, inception-resnet and the impact of
  residual connections on learning}, in: \bibinfo{booktitle}{Thirty-first AAAI
  conference on artificial intelligence}.
\bibitem[{Szegedy et~al.(2016)Szegedy, Vanhoucke, Ioffe, Shlens and
  Wojna}]{Szegedy2016}
\bibinfo{author}{Szegedy, C.}, \bibinfo{author}{Vanhoucke, V.},
  \bibinfo{author}{Ioffe, S.}, \bibinfo{author}{Shlens, J.},
  \bibinfo{author}{Wojna, Z.}, \bibinfo{year}{2016}.
\newblock \bibinfo{title}{Rethinking the inception architecture for computer
  vision}, in: \bibinfo{booktitle}{The IEEE Conference on Computer Vision and
  Pattern Recognition (CVPR)}, pp. \bibinfo{pages}{2818--2826}.
\bibitem[{Tan and Le(2019)}]{Tan2019}
\bibinfo{author}{Tan, M.}, \bibinfo{author}{Le, Q.}, \bibinfo{year}{2019}.
\newblock \bibinfo{title}{{E}fficient{N}et: Rethinking model scaling for
  convolutional neural networks}, in: \bibinfo{editor}{Chaudhuri, K.},
  \bibinfo{editor}{Salakhutdinov, R.} (Eds.), \bibinfo{booktitle}{Proceedings
  of the 36th International Conference on Machine Learning},
  \bibinfo{publisher}{PMLR}. pp. \bibinfo{pages}{6105--6114}.
\newblock \URLprefix \url{https://proceedings.mlr.press/v97/tan19a.html}.
\bibitem[{Tsai et~al.(2021)Tsai, Simpson, Lungren, Hershman, Roshkovan, Colak,
  Erickson, Shih, Stein, Kalpathy-Cramer et~al.}]{Tsai2021}
\bibinfo{author}{Tsai, E.}, \bibinfo{author}{Simpson, S.},
  \bibinfo{author}{Lungren, M.}, \bibinfo{author}{Hershman, M.},
  \bibinfo{author}{Roshkovan, L.}, \bibinfo{author}{Colak, E.},
  \bibinfo{author}{Erickson, B.}, \bibinfo{author}{Shih, G.},
  \bibinfo{author}{Stein, A.}, \bibinfo{author}{Kalpathy-Cramer, J.}, et~al.,
  \bibinfo{year}{2021}.
\newblock \bibinfo{title}{‘data from medical imaging data resource center
  (midrc)-rsna international covid radiology database (ricord) release
  1c—chest x-ray, covid+(midrc-ricord-1c)}.
\newblock \bibinfo{journal}{The Cancer Imaging Archive. DOI: https://doi.
  org/10.7937/91ah-v663} .
\bibitem[{Wang et~al.(2017)Wang, Peng, Lu, Lu, Bagheri and Summers}]{Wang2017}
\bibinfo{author}{Wang, X.}, \bibinfo{author}{Peng, Y.}, \bibinfo{author}{Lu,
  L.}, \bibinfo{author}{Lu, Z.}, \bibinfo{author}{Bagheri, M.},
  \bibinfo{author}{Summers, R.M.}, \bibinfo{year}{2017}.
\newblock \bibinfo{title}{Chestx-ray8: Hospital-scale chest x-ray database and
  benchmarks on weakly-supervised classification and localization of common
  thorax diseases}, in: \bibinfo{booktitle}{Proceedings of the IEEE conference
  on computer vision and pattern recognition}, pp. \bibinfo{pages}{2097--2106}.
\bibitem[{{World Health Organization}(2022)}]{WHO2022}
\bibinfo{author}{{World Health Organization}}, \bibinfo{year}{2022}.
\newblock \bibinfo{title}{Who coronavirus (covid-19) dashboard}.
\newblock \URLprefix \url{https://covid19.who.int/}. \bibinfo{note}{accessed:
  2022-02-28}.
\bibitem[{Zhao et~al.(2021)Zhao, Jiang and Qiu}]{Zhao2021}
\bibinfo{author}{Zhao, W.}, \bibinfo{author}{Jiang, W.}, \bibinfo{author}{Qiu,
  X.}, \bibinfo{year}{2021}.
\newblock \bibinfo{title}{Fine-tuning convolutional neural networks for
  covid-19 detection from chest x-ray images}.
\newblock \bibinfo{journal}{Diagnostics} \bibinfo{volume}{11},
  \bibinfo{pages}{1887}.
\newblock \URLprefix \url{http://dx.doi.org/10.3390/diagnostics11101887},
  \DOIprefix\doi{10.3390/diagnostics11101887}.
\bibitem[{Zoph et~al.(2018)Zoph, Vasudevan, Shlens and Le}]{Zoph2018}
\bibinfo{author}{Zoph, B.}, \bibinfo{author}{Vasudevan, V.},
  \bibinfo{author}{Shlens, J.}, \bibinfo{author}{Le, Q.V.},
  \bibinfo{year}{2018}.
\newblock \bibinfo{title}{Learning transferable architectures for scalable
  image recognition}, in: \bibinfo{booktitle}{The IEEE Conference on Computer
  Vision and Pattern Recognition (CVPR)}, pp. \bibinfo{pages}{8697--8710}.

\end{thebibliography}

\appendix

\section{Classification results in training, validation, and test subsets}
\label{sec:AppendixSubsets}

This appendix shows the classification results obtained by the individual CNN classifiers, with the same weights learned in the Section~\ref{sec:CNNComparison} experiment when applied to the training, validation, and test subsets individually. All results are the average of the five executions, with different training/validation splits. Tables~\ref{tab:SubsetsACC}, \ref{tab:SubsetsTPR}, \ref{tab:SubsetsPPV}, \ref{tab:SubsetsF1} show the results of accuracy (ACC), sensitivity (TPR), precision (PPV), and F1 score, respectively.

\begin{table}
\centering

\begin{tabular}{rccc}

\toprule

{\bf Dataset / Subset} & {\bf Train} & {\bf Validation} & {\bf Test} \\

\toprule

DenseNet169 &     0.9951 &     0.9794 &     0.9815 \\

EfficientNetB2 &     0.9936 &     0.9793 &     0.9760 \\

InceptionResNetV2 &     0.9835 &     0.9681 &     0.9755 \\

InceptionV3 &     0.9960 &     0.9784 &     0.9750 \\

 MobileNet &     0.9936 &     0.9788 &     0.9710 \\

EfficientNetB0 &     0.9894 &     0.9761 &     0.9705 \\

EfficientNetB3 &     0.9948 &     0.9803 &     0.9700 \\

ResNet152V2 &     0.9945 &     0.9757 &     0.9695 \\

DenseNet201 &     0.9971 &     0.9816 &     0.9695 \\

 ResNet152 &     0.9923 &     0.9783 &     0.9660 \\

DenseNet121 &     0.9962 &     0.9806 &     0.9630 \\

  Xception &     0.9909 &     0.9777 &     0.9615 \\

     VGG19 &     0.9922 &     0.9804 &     0.9580 \\

EfficientNetB1 &     0.9802 &     0.9697 &     0.9570 \\

  ResNet50 &     0.9955 &     0.9806 &     0.9545 \\

ResNet101V2 &     0.9909 &     0.9707 &     0.9530 \\

     VGG16 &     0.9913 &     0.9772 &     0.9525 \\

MobileNetV2 &     0.9987 &     0.9808 &     0.9485 \\

 ResNet101 &     0.9923 &     0.9803 &     0.9410 \\

ResNet50V2 &     0.9859 &     0.9662 &     0.9280 \\

NASNetMobile &     0.9798 &     0.9660 &     0.8530 \\

\bottomrule

\end{tabular}

\caption{Classification accuracy (ACC) achieved by the CNN architectures when applied to the train, validation, and test subsets individually.}
\label{tab:SubsetsACC}
\end{table}

\begin{table}
\centering

\begin{tabular}{rccc}

\toprule

{\bf Dataset / Subset} & {\bf Train} & {\bf Validation} & {\bf Test} \\

\toprule

DenseNet169 &     0.9987 &     0.9611 &     0.9700 \\

EfficientNetB2 &     0.9936 &     0.9662 &     0.9600 \\

InceptionResNetV2 &     0.9830 &     0.9491 &     0.9590 \\

InceptionV3 &     0.9965 &     0.9458 &     0.9520 \\

EfficientNetB0 &     0.9760 &     0.9361 &     0.9510 \\

EfficientNetB3 &     0.9935 &     0.9662 &     0.9470 \\

 MobileNet &     0.9957 &     0.9440 &     0.9430 \\

ResNet152V2 &     0.9928 &     0.9375 &     0.9420 \\

DenseNet201 &     0.9964 &     0.9454 &     0.9400 \\

 ResNet152 &     0.9854 &     0.9509 &     0.9370 \\

DenseNet121 &     0.9973 &     0.9421 &     0.9270 \\

EfficientNetB1 &     0.9750 &     0.9505 &     0.9240 \\

  Xception &     0.9847 &     0.9333 &     0.9230 \\

     VGG19 &     0.9874 &     0.9338 &     0.9170 \\

ResNet101V2 &     0.9882 &     0.9278 &     0.9100 \\

     VGG16 &     0.9915 &     0.9324 &     0.9090 \\

  ResNet50 &     0.9918 &     0.9338 &     0.9090 \\

MobileNetV2 &     0.9933 &     0.9241 &     0.9030 \\

 ResNet101 &     0.9701 &     0.9162 &     0.8830 \\

ResNet50V2 &     0.9758 &     0.8921 &     0.8590 \\

NASNetMobile &     0.8874 &     0.8255 &     0.7090 \\

\bottomrule

\end{tabular}

\caption{Classification sensitivity (TPR) achieved by the CNN architectures when applied to the train, validation, and test subsets individually.}
\label{tab:SubsetsTPR}
\end{table}

\begin{table}
\centering

\begin{tabular}{rccc}

\toprule

{\bf Dataset / Subset} & {\bf Train} & {\bf Validation} & {\bf Test} \\

\toprule

  Xception &     0.9502 &     0.9057 &     1.0000 \\

  ResNet50 &     0.9759 &     0.9242 &     1.0000 \\

 MobileNet &     0.9587 &     0.9040 &     0.9990 \\

     VGG19 &     0.9566 &     0.9228 &     0.9989 \\

DenseNet121 &     0.9754 &     0.9169 &     0.9989 \\

DenseNet201 &     0.9822 &     0.9215 &     0.9989 \\

 ResNet101 &     0.9727 &     0.9366 &     0.9988 \\

InceptionV3 &     0.9748 &     0.8998 &     0.9979 \\

ResNet152V2 &     0.9682 &     0.8904 &     0.9970 \\

ResNet50V2 &     0.9244 &     0.8630 &     0.9966 \\

NASNetMobile &     0.9602 &     0.9159 &     0.9960 \\

ResNet101V2 &     0.9501 &     0.8701 &     0.9959 \\

     VGG16 &     0.9474 &     0.9044 &     0.9958 \\

 ResNet152 &     0.9598 &     0.8964 &     0.9947 \\

MobileNetV2 &     0.9973 &     0.9337 &     0.9935 \\

DenseNet169 &     0.9672 &     0.8946 &     0.9930 \\

EfficientNetB3 &     0.9702 &     0.8977 &     0.9927 \\

InceptionResNetV2 &     0.9077 &     0.8408 &     0.9919 \\

EfficientNetB2 &     0.9631 &     0.8925 &     0.9918 \\

EfficientNetB0 &     0.9486 &     0.8928 &     0.9896 \\

EfficientNetB1 &     0.8924 &     0.8470 &     0.9892 \\

\bottomrule

\end{tabular}

\caption{Classification precision (PPV) achieved by the CNN architectures when applied to the train, validation, and test subsets individually.}
\label{tab:SubsetsPPV}
\end{table}

\begin{table}
\centering

\begin{tabular}{rccc}

\toprule

{\bf Dataset / Subset} & {\bf Train} & {\bf Validation} & {\bf Test} \\

\toprule

DenseNet169 &     0.9825 &     0.9266 &     0.9812 \\

EfficientNetB2 &     0.9774 &     0.9273 &     0.9756 \\

InceptionResNetV2 &     0.9428 &     0.8905 &     0.9749 \\

InceptionV3 &     0.9855 &     0.9221 &     0.9744 \\

 MobileNet &     0.9768 &     0.9233 &     0.9701 \\

EfficientNetB0 &     0.9619 &     0.9139 &     0.9699 \\

EfficientNetB3 &     0.9815 &     0.9304 &     0.9690 \\

DenseNet201 &     0.9892 &     0.9331 &     0.9683 \\

ResNet152V2 &     0.9801 &     0.9128 &     0.9679 \\

 ResNet152 &     0.9722 &     0.9226 &     0.9644 \\

DenseNet121 &     0.9862 &     0.9292 &     0.9616 \\

  Xception &     0.9670 &     0.9190 &     0.9599 \\

     VGG19 &     0.9716 &     0.9281 &     0.9558 \\

EfficientNetB1 &     0.9312 &     0.8954 &     0.9551 \\

  ResNet50 &     0.9837 &     0.9287 &     0.9520 \\

     VGG16 &     0.9688 &     0.9173 &     0.9501 \\

ResNet101V2 &     0.9679 &     0.8966 &     0.9497 \\

MobileNetV2 &     0.9953 &     0.9287 &     0.9457 \\

 ResNet101 &     0.9714 &     0.9262 &     0.9370 \\

ResNet50V2 &     0.9493 &     0.8773 &     0.9226 \\

NASNetMobile &     0.9209 &     0.8675 &     0.8212 \\

\bottomrule

\end{tabular}

\caption{Classification F1 score achieved by the CNN architectures when applied to the train, validation, and test subsets individually.}
\label{tab:SubsetsF1}
\end{table}

These results show that accuracy and sensitivity are a little higher in the training and validation subsets than in the test subset for all the architectures. On the other hand, precision is higher on the test subset for all the architectures. Finally, the F1 score shows the closest results in the training and test subsets. Some architectures achieved a higher F1 score in the training subset, while others achieved a higher F1 score in the test subset. This behavior may be related to the class imbalance, even though class weights were used to minimize its effects. Future work may focus on other techniques to handle class imbalance, like data augmentation. Moreover, future benchmark datasets will probably minimize imbalance as more COVID-19 CXR images become available.

\end{document}